\documentclass[nolinenumbers]{aastex631}
\makeatletter
\let\frontmatter@title@above=\relax
\makeatother

\usepackage{xspace}
\usepackage[figure,figure*]{hypcap}
\usepackage{multirow}
\usepackage{hyperref}
\usepackage{xurl}

\newcommand{\OIII}{[\textrm{O}~\textsc{iii}]\xspace}

\newcommand{\grizli}{\textsc{Grizli}}
\newcommand{\eazy}{\textsc{Eazy}}
\newcommand{\sep}{\textsc{SEP}}

\received{November 22, 2024}
\accepted{August 25, 2025}

\shorttitle{PASSAGE Overview} 

 \shortauthors{Malkan et al.}

\begin{document}
\title{Parallel Application of Slitless Spectroscopy to Analyze Galaxy Evolution (PASSAGE): Survey Overview }

\author[0000-0001-6919-1237]{Matthew A. Malkan}
\affiliation{University of California, Los Angeles, Department of Physics and Astronomy, 430 Portola Plaza, Los Angeles, CA 90095, USA}

\author[0000-0001-7166-6035]{Vihang Mehta}
\affiliation{IPAC, California Institute of Technology, 1200 E. California Blvd, Pasadena, CA 91125, USA}

\author[0000-0003-4804-7142]{Ayan Acharyya}
\affiliation{INAF, Osservatorio Astronomico di Padova, Vicolo dell'Osservatorio 5, 35122 Padova, Italy}

\author[0000-0003-3596-8794]{Hollis B. Akins}
\affiliation{The University of Texas at Austin, 2515 Speedway Blvd Stop C1400, Austin, TX 78712, USA}

\author[0000-0002-8630-6435]{Anahita Alavi}
\affiliation{IPAC, California Institute of Technology, 1200 E. California Blvd, Pasadena, CA 91125, USA}

\author[0000-0002-7570-0824]{Hakim Atek}
\affiliation{CNRS, Institut d'Astrophysique de Paris, 98 bis Boulevard Arago, 75014 Paris, France}

\author[0000-0003-0556-2929]{Ivano Baronchelli}
\affiliation{INAF, Istituto di Radioastronomia, Via Piero Gobetti 101, 40129 Bologna, Italy}

\author[0000-0003-4569-2285]{Andrew J. Battisti}
\affiliation{International Centre for Radio Astronomy Research (ICRAR), University of Western Australia, M468, 35 Stirling Highway, Crawley, WA 6009, Australia}
\affiliation{Australian National University, Research School of Astronomy and Astrophysics, Canberra, ACT 2611, Australia}
\affiliation{ARC Centre of Excellence for All Sky Astrophysics in 3 Dimensions (ASTRO 3D), Australia}

\author[0000-0003-4109-304X]{Kit Boyett}
\affiliation{School of Physics, University of Melbourne, Parkville 3010, VIC, Australia}
\affiliation{ARC Centre of Excellence for All Sky Astrophysics in 3 Dimensions (ASTRO 3D), Australia}
\affiliation{University of Oxford, Department of Physics, Keble Road, Oxford OX1 3RH, UK}

\author[0000-0001-5984-0395]{Marusa Bradac}
\affiliation{University of Ljubljana, Department of Physics, Jadranska ulica 19, 1000 Ljubljana, Slovenia}

\author[0000-0002-6503-5218]{Sean Tyler Bruton}
\affiliation{University of Minnesota, Twin Cities, 116 Church St SE, Minneapolis, MN 55455, USA}

\author[0000-0002-8651-9879]{Andrew J.\ Bunker}
\affiliation{University of Oxford, Department of Physics, Keble Road, Oxford OX1 3RH, UK}

\author[0000-0002-6523-9536]{Adam J. Burgasser}
\affiliation{University of California, San Diego, Center for Astrophysics and Space Sciences, 9500 Gilman Drive, La Jolla, CA 92093, USA}

\author[0000-0002-0930-6466]{Caitlin M. Casey}
\affiliation{Department of Physics, University of California, Santa Barbara, Santa Barbara, CA 93106, USA}
\affiliation{The University of Texas at Austin, 2515 Speedway Blvd Stop C1400, Austin, TX 78712, USA}
\affiliation{Cosmic Dawn Center (DAWN), Denmark}

\author[0000-0002-0486-5242]{Nuo Chen}
\affiliation{The University of Tokyo, Graduate School of Science, Department of Astronomy, 7-3-1 Hongo, Bunkyo-ku, Tokyo 113-0033, Japan}

\author[0000-0001-6482-3020]{James Colbert}
\affiliation{IPAC, California Institute of Technology, 1200 E. California Blvd, Pasadena, CA 91125, USA}

\author[0000-0002-7928-416X]{Y. Sophia Dai}
\affiliation{The National Astronomical Observatories, Chinese Academy of Sciences, 20A Datun Road, Chaoyang District, Beijing 100101, China}

\author[0000-0002-3560-8599]{Maximilien Franco}
\affiliation{Université Paris-Saclay, Université Paris Cité, CEA, CNRS, AIM, 91191 Gif-sur-Yvette, France}
\affiliation{The University of Texas at Austin, 2515 Speedway Blvd Stop C1400, Austin, TX 78712, USA}

\author[0009-0009-0868-8165]{Clea Hannahs}
\affiliation{University of California, Los Angeles, Department of Physics and Astronomy, 430 Portola Plaza, Los Angeles, CA 90095, USA}

\author[0000-0003-0129-2079]{Santosh Harish}
\affiliation{Laboratory for Multiwavelength Astrophysics, School of Physics and Astronomy, Rochester Institute of Technology, 84 Lomb Memorial Drive, Rochester, NY 14623, USA}

\author[0000-0002-0072-0281]{Farhanul Hasan}
\affiliation{Space Telescope Science Institute, 3700 San Martin Drive, Baltimore, MD 21218, USA}

\author[0000-0001-8587-218X]{Matthew J. Hayes}
\affiliation{Stockholm University, Department of Astronomy, AlbaNova University Center, SE-106 91 Stockholm, Sweden}

\author[0000-0002-6586-4446]{Alaina L. Henry}
\affiliation{Space Telescope Science Institute, 3700 San Martin Drive, Baltimore, MD 21218, USA}

\author[0009-0002-9932-4461]{Mason Huberty}
\affiliation{University of Minnesota, Twin Cities, 116 Church St SE, Minneapolis, MN 55455, USA}

\author[0000-0001-9187-3605]{Jeyhan S. Kartaltepe}
\affiliation{Laboratory for Multiwavelength Astrophysics, School of Physics and Astronomy, Rochester Institute of Technology, 84 Lomb Memorial Drive, Rochester, NY 14623, USA}

\author[0000-0001-6505-0293]{Keunho J. Kim}
\affiliation{IPAC, California Institute of Technology, 1200 E. California Blvd, Pasadena, CA 91125, USA}

\author[0000-0003-4570-3159]{Nicha Leethochawalit}
\affiliation{National Astronomical Research Institute of Thailand (NARIT), Mae Rim, Chiang Mai, 50180, Thailand}

\author[0009-0000-9478-1933]{Jacob Levine}
\affiliation{University of California, Los Angeles, Department of Physics and Astronomy, 430 Portola Plaza, Los Angeles, CA 90095, USA}

\author[0000-0003-4813-8482]{Sijia Li}
\affil{School of Astronomy and Space Science, University of Chinese Academy of Sciences (UCAS), Beijing 100049, China}
\affil{Department of Astronomy, Xiamen University, Xiamen, Fujian 361005, China}

\author[0000-0001-8792-3091]{Yu-Heng Lin}
\affiliation{IPAC, California Institute of Technology, 1200 E. California Blvd, Pasadena, CA 91125, USA}

\author[0009-0006-4990-7529]{Yixiao Liu}
\affiliation{The National Astronomical Observatories, Chinese Academy of Sciences, 20A Datun Road, Chaoyang District, Beijing 100101, China}

\author[0000-0002-3407-1785]{Charlotte Mason}
\affiliation{University of Copenhagen, Niels Bohr Institute, Juliane Maries Vej 30, DK-2100 Copenhagen, Denmark}

\author[0000-0001-5382-6138]{Daniel Masters}
\affiliation{The Johns Hopkins University, Department of Physics and Astronomy, 3400 N. Charles Street, Baltimore, MD 21218, USA}

\author[0000-0002-9489-7765]{Henry Joy McCracken}
\affiliation{Institut d’Astrophysique de Paris, UMR 7095, CNRS, and Sorbonne Université, 98 bis boulevard Arago, F-75014 Paris, France}

\author[0000-0002-8512-1404]{Takahiro Morishita}
\affiliation{IPAC, California Institute of Technology, 1200 E. California Blvd, Pasadena, CA 91125, USA}

\author[0000-0001-5294-8002]{Kalina V. Nedkova}
\affiliation{The Johns Hopkins University, Department of Physics and Astronomy, 3400 N. Charles Street, Baltimore, MD 21218, USA}
\affiliation{Space Telescope Science Institute, 3700 San Martin Drive, Baltimore, MD 21218, USA}

\author[0000-0002-9946-4731]{Marc Rafelski}
\affiliation{Space Telescope Science Institute, 3700 San Martin Drive, Baltimore, MD 21218, USA}
\affiliation{The Johns Hopkins University, Department of Physics and Astronomy, 3400 N. Charles Street, Baltimore, MD 21218, USA}

\author{Vivasvaan Aditya Raj}
\affiliation{University of California, Los Angeles, Department of Physics and Astronomy, 430 Portola Plaza, Los Angeles, CA 90095, USA}

\author[0000-0002-4140-1367]{Guido Roberts-Borsani}
\affiliation{Department of Astronomy, University of Geneva, 51 Chemin Pegasi, CH-1290 Versoix, Switzerland}

\author[0000-0002-1025-7569]{Axel Runnholm}
\affiliation{Stockholm University, Department of Astronomy, AlbaNova University Center, SE-106 91 Stockholm, Sweden}

\author[0000-0001-7016-5220]{Michael J. Rutkowski}
\affiliation{Minnesota State University, Mankato, Department of Physics and Astronomy, 141 Trafton Science Center N, Mankato, MN 56001, USA}

\author[0000-0001-8419-3062]{Alberto Saldana-Lopez}
\affiliation{Department of Astronomy, Oskar Klein Centre, Stockholm University, 106 91 Stockholm, Sweden}

\author[0000-0002-0364-1159]{Zahra Sattari}
\affiliation{IPAC, California Institute of Technology, 1200 E. California Blvd, Pasadena, CA 91125, USA}

\author[0000-0002-9136-8876]{Claudia Scarlata}
\affiliation{University of Minnesota, Twin Cities, 116 Church St SE, Minneapolis, MN 55455, USA}

\author[0000-0002-3418-7251]{Kasper Borello Schmidt}
\affiliation{University of California, Los Angeles, Department of Physics and Astronomy, 430 Portola Plaza, Los Angeles, CA 90095, USA}

\author[0000-0002-7087-0701]{Marko Shuntov}
\affiliation{Cosmic Dawn Center (DAWN), Denmark} 
\affiliation{Niels Bohr Institute, University of Copenhagen, Jagtvej 128, DK-2200, Copenhagen, Denmark}

\author[0000-0002-7064-5424]{Harry Teplitz}
\affiliation{California Institute of Technology, 1200 E. California Blvd, Pasadena, CA 91125, USA}

\author[0000-0001-9391-305X]{Michele Trenti}
\affiliation{School of Physics, University of Melbourne, Parkville 3010, VIC, Australia}
\affiliation{ARC Centre of Excellence for All Sky Astrophysics in 3 Dimensions (ASTRO 3D), Australia}

\author[0000-0002-8460-0390]{Tommaso Treu}
\affiliation{University of California, Los Angeles, Department of Physics and Astronomy, 430 Portola Plaza, Los Angeles, CA 90095, USA}

\author[0000-0003-0980-1499]{Benedetta Vulcani}
\affiliation{INAF, Osservatorio Astronomico di Padova, Vicolo dell'Osservatorio 5, 35122 Padova, Italy}

\author[0000-0003-3108-0624]{Peter J. Watson}
\affiliation{INAF, Osservatorio Astronomico di Padova, Vicolo dell'Osservatorio 5, 35122 Padova, Italy}

\author[0000-0002-9373-3865]{Xin Wang}
\affiliation{School of Astronomy and Space Science, University of Chinese Academy of Sciences (UCAS), Beijing 100049, China}
\affiliation{National Astronomical Observatories, Chinese Academy of Sciences, Beijing 100101, China}
\affiliation{Institute for Frontiers in Astronomy and Astrophysics, Beijing Normal University, Beijing 102206, China}

\author[0000-0002-1945-2299]{Zhuyun Zhuang}
\affiliation{California Institute of Technology, 1200 E. California Blvd, Pasadena, CA 91125, USA}

\correspondingauthor{Matthew Malkan}
\email{malkan@astro.ucla.edu}





\begin{abstract}
During the second half of Cycle 1 of the James Webb Space Telescope (JWST), we conducted the Parallel Application of Slitless Spectroscopy to Analyze Galaxy Evolution (PASSAGE) program. PASSAGE received the largest  allocation of JWST observing time in Cycle 1, 591 hours of NIRISS observations to obtain direct near-IR imaging and slitless spectroscopy. About two thirds of these were ultimately executed, to observe 63 high-latitude fields in Pure Parallel mode. These have provided more than ten thousand near-infrared grism spectrograms of faint galaxies.

PASSAGE brings unique advantages in studying galaxy evolution:
A) Unbiased spectroscopic search, without prior photometric pre-selection. By including the most numerous galaxies, with low masses and strong emission lines, slitless spectroscopy is the indispensable complement to any pre-targeted spectroscopy; 
B) \noindent The combination of several dozen independent fields to overcome cosmic variance;  
C) \noindent Near-infrared spectral coverage, often spanning the full range from 1.0--2.3 $\mu$m, with minimal wavelength gaps, to measure multiple diagnostic rest-frame optical lines, minimizing sensitivity to dust reddening; 
D) \noindent JWST's unprecedented spatial resolution, in some cases using two orthogonal grism orientations, to overcome contamination due to blending of overlapping spectra;
E) Discovery of rare bright objects especially for detailed JWST followup. PASSAGE data are public immediately, and our team plans to deliver fully-processed high-level data products.

In this PASSAGE overview, we describe the survey and data quality, and present examples of these accomplishments in several areas of current interest in the evolution of emission-line galaxy properties, particularly at low masses.

\end{abstract}


\section{Introduction}

The value of unbiased spectroscopic selections of galaxies has long been recognized for understanding their evolution across cosmic time. For emission-line galaxies, this can crudely be done with ground-based imaging through narrow-band filters
\citep{1999ApJ...514...33T, 2006ApJ...648....7K, 2007ApJ...657..738L, 2011ApJ...734..119K, 2012ApJ...757...63L, 2018ApJ...863...92B, bunker1995}.
Full spectral surveys have been done using
grism spectroscopy with HST/NICMOS,  \citep{1999ApJ...520..548M, 1999ApJ...519L..47Y, 2002ApJ...581..205H, 2024ApJ...962..195B}, and then with HST/WFC3 \citep{2010ApJ...723..104A, 2013ASPC..477..255M, Henry2013, 2017ApJ...837...11B, Henry2021, 2024MNRAS.530..894B, 2015ApJ...812..114T}. More recently,
JWST has also used grisms to obtain slitless spectroscopy of large samples of galaxies, e.g. \citep{Estrada2024, He2024, Bradac2024, Rihtarv2025, Watson2025}.

\vskip 0.10 in
Cosmic star formation likely peaked at $z\sim$ 2--3 \citep{Ly2009, Madau2014}.
Most of our knowledge of galaxies at these epochs is restricted to those which are massive. 
We know less about the more representative galaxies with  $\log(M_\ast/M_\odot)\leq10$, as samples are sparse and biased, with the stacking of low-SNR spectra often required 
\citep[e.g.,][]{Henry2013, Sanders2018, Henry2021}.
Parallel NIRISS slitless spectroscopy is revolutionizing our
understanding of galaxy evolution at high redshifts, especially in the important low-mass regime. By combining many fields with moderate depth, plus smaller numbers of deep fields, PASSAGE obtains coverage of statistically powerful numbers of galaxies spanning a very wide range of luminosities and stellar masses. We are measuring key optical emission line diagnostics, including [OII]$\lambda$3726,3729\AA, [OIII]4959,5007\AA, and hydrogen Balmer lines, across a wide range of redshift ($z=1-3.5$), and Paschen and [SIII]9532,9069 lines at lower redshifts, for of order ten thousand galaxies--by far the largest sample ever assembled. 

In this paper we discuss our new PASSAGE observations in Section \ref{sec:Observations}, the data reduction and sensitivity in Section \ref{sec:Data Reduction}, some examples of scientific results in Section \ref{sec:faint Emission Line}, and future prospects in Section \ref{sec:Summary}.

\section{Observations} \label{sec:Observations}
\subsection{Slitless Spectroscopy with NIRISS}

The Near Infrared Imager and Slitless Spectrograph (NIRISS) instrument on the James Webb Space Telescope (JWST) enables near-infrared slitless spectroscopy (0.8--2.3~\micron) with two orthogonal medium resolution ($R\sim150$ at 1.4~\micron) grisms over a 133\arcsec$\times$133\arcsec\ field of view \citep{Willott_2022,Doyon_2023}. NIRISS employs a set of specialized broad filters (PASSAGE uses F115W, F150W, and F200W), to selectively transmit specific wavelength ranges of infrared light, crucial for isolating key spectral features while blocking out-of-band radiation. The filters work in conjunction with the NIRISS grisms (GR150R and GR150C), enabling slitless spectroscopy by dispersing light across the detector, along either rows or columns, respectively. The combination of these filters and grisms helps minimize contamination from overlapping spectral orders and maximizes sensitivity in the target wavelength range, making them essential for capturing detailed spectrograms of faint astronomical sources. By using multiple filters in sequence, NIRISS can construct a comprehensive spectral profile of objects. 

For the first Cycle of  (JWST Cy1), a large program of pure parallel spectroscopy received an allocation of 591 hours. 
PASSAGE, JWST-1571 (PI: Malkan)--began in December 2022, when the pure parallel observing mode first became available for scheduling. Pure-parallel observations, as implied, are executed simultaneously with a separate instrument (NIRISS in our case) in parallel while a primary instrument is being used by a different guest observer program. JWST pure-parallel observations are subject to various constraints imposed by the prime observation. A collection of prime observations were assigned to our program where we were allowed to craft an observing strategy for the NIRISS pure-parallel observations -- we refer to these as \textit{parallel opportunities}.

Scheduling pure parallel slitless spectroscopy is challenging.  Ideally, each wavelength setting requires 3 filter wheel motions (for a direct image and spectra with both orthogonal orientations of the dispersion axis).  However, to eliminate any risk of losing telescope guiding, these changes can only be made when the primary observers are simultaneously reconfiguring their instrument, and not taking data. In many parallel opportunities, this restricts the number of wavelengths that can be covered, especially when the primary observer is mosaicking more than one sky region.

We individually crafted observations of each available pure parallel field to optimize the use of every parallel opportunity, down to 1.5 hours in length. These PASSAGE observations were adapted to any combination of primary exposure  times and filter changes.
Since the primary objective of the PASSAGE survey was to maximize the wavelength range of the spectroscopy, in opportunities with more allowed filter changes, our first priority was to extend wavelength coverage by adding spectroscopy through more broadband blocking filters (F115W, F150W, and F200W). 
The secondary objective when defining the observations was to reduce contamination due to overlapping spectra, and hence, our next priority was to add observations with the second GR150 grism (usually GR150C after GR150R), providing spectrograms at a second, orthogonal, orientation.
We excluded potential fields at Galactic latitudes below 20 deg, to avoid spectrum overlaps in crowded fields. 

\begin{figure}
    \plotone{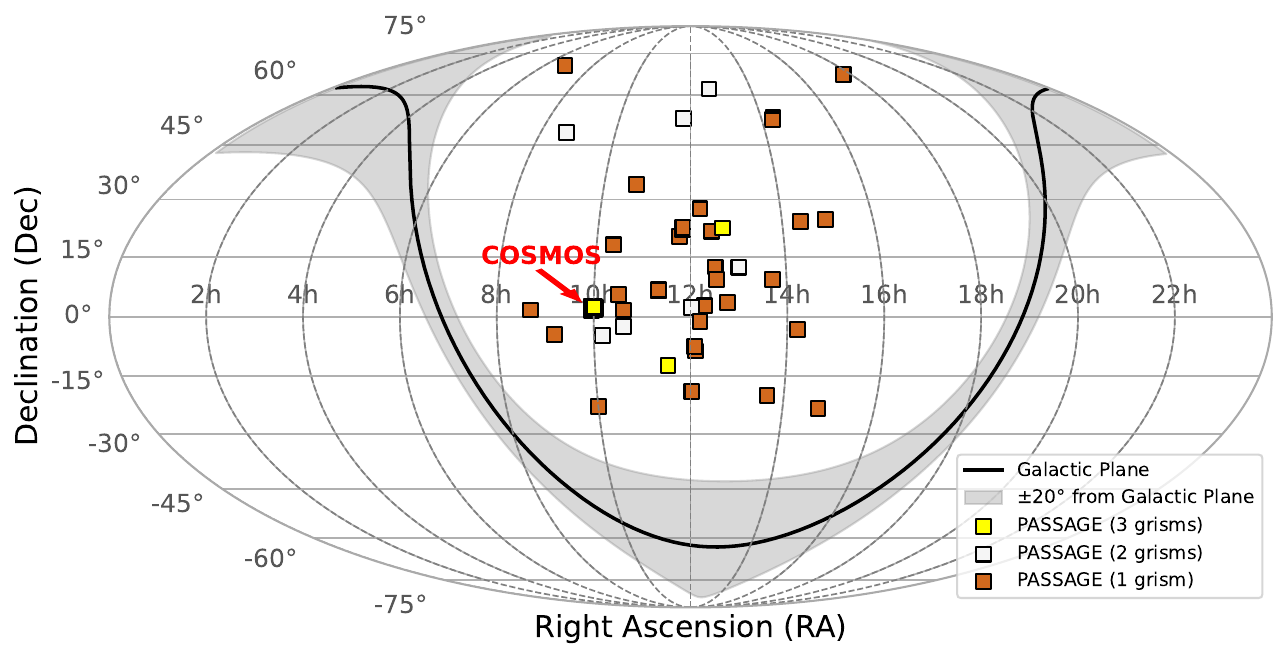}
    \caption{Sky locations of the final observed 63 PASSAGE fields, shown in Equatorial coordinates. Note that some fields overlap on the scale of this diagram. In particular, the 18 PASSAGE fields in the COSMOS field all overlap in a single yellow box in this low-resolution sky map. These are resolved in the COSMOS region zoom shown in Figure \ref{fig:cosmosoverlap}.}
    \label{fig:allsky}
\end{figure}

\begin{figure}
    \plotone{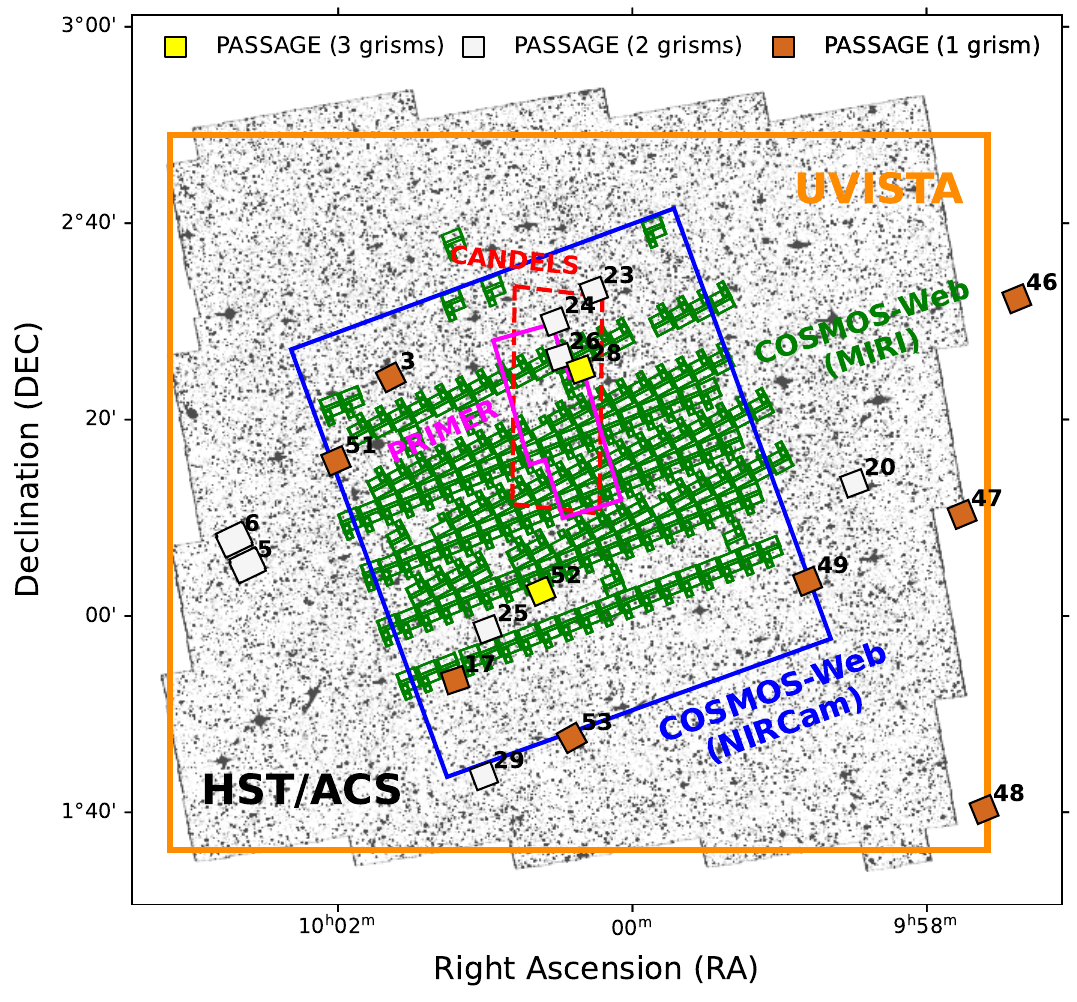}
    \caption{Zoom-in on PASSAGE fields located in and around the COSMOS field. The background is the Hubble ACS mosaic of the COSMOS field \citep{Scoville2007,Koekemoer2007}, along with the deep CANDELS survey as red outline \citep{Grogin2011,Koekemoer2011}. The orange outline shows the UltraVISTA imaging survey footprint of the COSMOS field \citep{McCracken2012}. The blue and green outline is the COSMOS-Web NIRCam and MIRI coverage, respectively \citep{Casey2023}. The magenta outline is the coverage of the PRIMER JWST survey (GO $\#$1837).}
    \label{fig:cosmosoverlap}
\end{figure}

Table \ref{table:observing_log} presents the PASSAGE Observing Log, detailing the sky coordinates of all NIRISS field centers observed to date, along with the exposure times for both direct imaging and spectroscopy in each waveband. While our exposure times were constrained by the scheduling of prime observations, we consistently assigned the longer integrations to spectroscopy, ensuring that the grism exposures have equal or greater duration than the imaging. The sky locations of these observations are illustrated in Figures \ref{fig:allsky} and \ref{fig:cosmosoverlap}.

\begin{figure}
    \centering
    \includegraphics[width=0.8\linewidth]{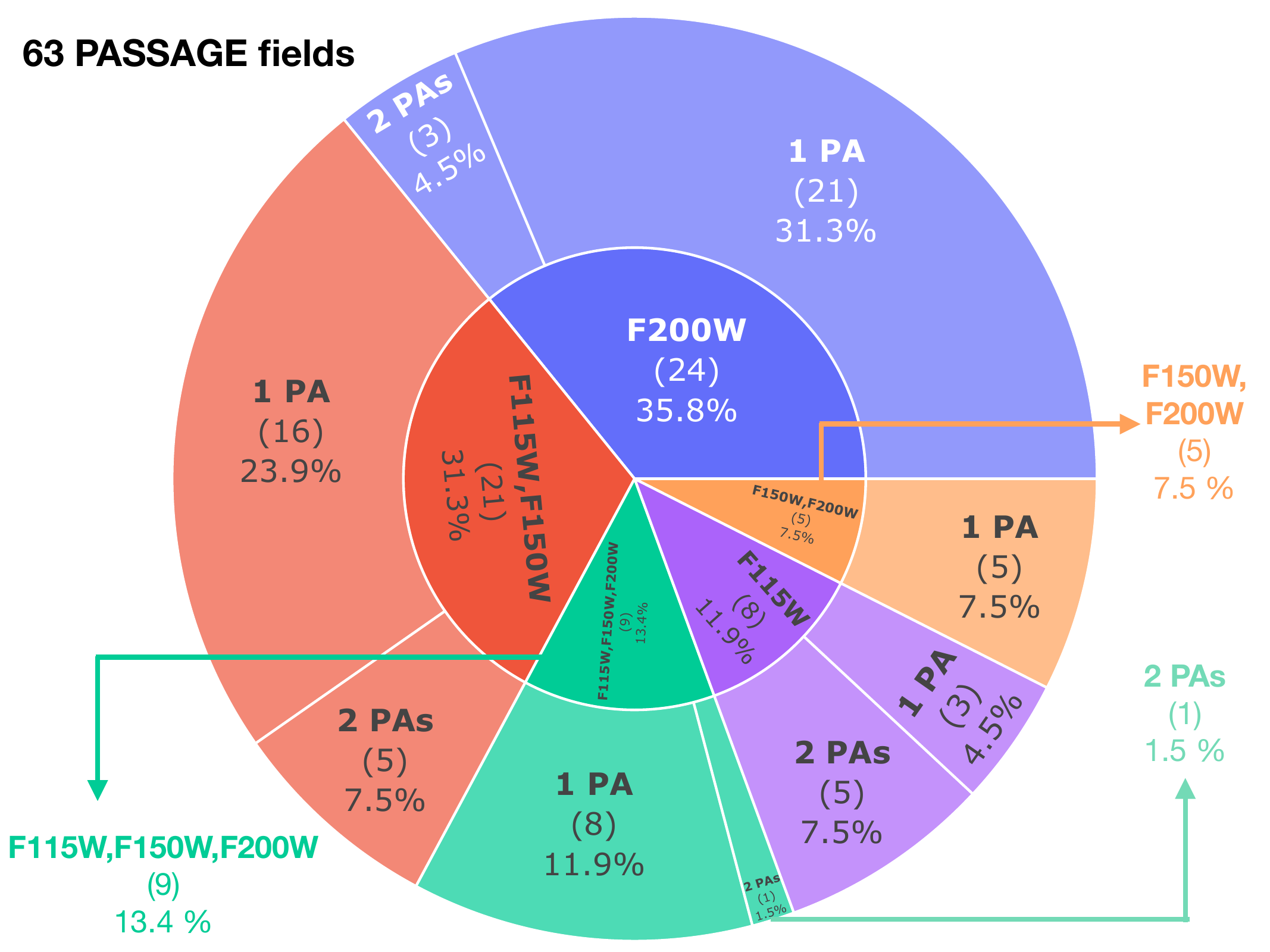}
    \caption{Pie chart showing the distribution of overlap percentages across different JWST filters (F115W, F150W, F200W) used in PASSAGE. The inner ring divides the observations into 1-, 2-, and 3-filter spectra.  The outer rings further separate these observations into those observed with a single grism (``1 PA", signifying that the spectra are dispersed at a single position angle), and those observed with both the GR150R and GR150C grisms.  The total number of fields in each category are given in parentheses. The total numbers in the wheel sum up to 67, rather than 63, due to  the 4 fields, such as Par027, which were observed with two different combinations of filters and orientations. To include all of the combinations used, these 4 fields were therefore double counted.}
    \label{fig:pie_chart}
\end{figure}

Because the fainter emission lines of interest (Ly$\alpha$ and [OII]) generally fall in the shorter-wavelength spectral coverage, in all medium- and long-exposure parallel fields, we typically integrated 2--2.5 times longer in that F115W broadband filter (1.0--1.3$\mu$m) than in the longer-wavelength filters. Most of the shorter parallels fields (up to 4 hours total) were devoted entirely to spectroscopy through the F200W broadband filter (1.7--2.3$\mu$m), because this long-wavelength region was inaccessible to slitless spectroscopy with WFC3 on HST \citep{2024MNRAS.530..894B}. The detailed breakdown of exposure combinations for the survey is illustrated in the pie-chart of Figure \ref{fig:pie_chart}.  Four out of the 63 PASSAGE fields have multiple combinations orientations and filters, leading to a double counting in the wheel of Figure \ref{fig:pie_chart}. For instance, a field that was observed with F115W in 2 PAs and with F150W and F200W with only 1 PA, contributes to both the purple and orange slices. Therefore, the number of fields in the wheels sum up to 67, rather than 63.

\begin{deluxetable*}{c|rrc|ccc|ccc}
\centering
\tablecaption{List of observed PASSAGE fields with their target coordinates and exposure times. \label{table:observing_log} }
\tabletypesize{\scriptsize}
\tablehead{
    \multirow{3}{*}{Field} &
    \multirow{3}{*}{R.A. [deg]} &
    \multirow{3}{*}{Decl [deg]} &
    \multirow{3}{*}{Obs Date} &
    F115W & F150W & F200W & F115W & F150W & F200W \\
    &&&& Direct & Direct & Direct & Grism & Grism & Grism \\
    &&&& [hr] & [hr] & [hr] & [hr] & [hr] & [hr] }
\startdata
    Par001 &    170.0267050 &      6.5888780 &     2022-12-17 & 0.43 & 0.86 &   -- & 2.30* & 4.60* &   --  \\
    Par002 &    177.1453720 &     52.7788100 &     2022-12-28 & 0.86 & 0.43 &   -- & 3.45* & 3.45* &   --  \\
    Par003 &    150.4103960 &      2.4051780 &     2022-12-30 &   -- &   -- & 0.62 &   -- &   -- & 1.24*  \\
    Par004 &    186.7047850 &     21.6989890 &     2022-12-31 &   -- &   -- & 1.15 &   -- &   -- & 1.15  \\
    Par005 &    150.6536980 &      2.0855800 &     2023-01-07 & 0.43 & 0.43 &   -- & 2.29* & 2.29* &   --  \\
    Par006 &    150.6762160 &      2.1293050 &     2023-01-06 & 0.43 & 0.43 &   -- & 2.29* & 2.29* &   --  \\
    Par007 &    102.4303090 &     70.3818910 &     2023-01-08 &   -- &   -- & 0.89 &   -- &   -- & 0.89  \\
    Par008 &    170.1264770 &      6.7062310 &     2023-01-12 &   -- &   -- & 0.86 &   -- &   -- & 1.72*  \\
    Par009 &    183.1322960 &     27.4870880 &     2023-01-13 &   -- &   -- & 0.52 &   -- &   -- & 1.19*  \\
    Par010 &    205.5720260 &      9.3080670 &     2023-02-21 & 0.17 &   -- &   -- & 3.01 & 1.00 &   --  \\
    Par011 &    204.5789780 &    -19.8794780 &     2023-02-22 & 2.61 &   -- &   -- & 2.61 &   -- &   --  \\
    Par012 &    221.6227940 &    -23.2860630 &     2023-02-25 &   -- &   -- & 0.01 &   -- &   -- & 1.91  \\
    Par013 &    130.5298840 &      1.6983950 &     2023-03-19 &   -- & 0.52 &   -- & 0.67 & 0.52 &   --  \\
    Par014 &    224.3278890 &     24.7169780 &     2023-03-23 & 0.67 &   -- &   -- & 0.67 &   -- &   --  \\
    Par015 &    155.4060630 &     18.1657440 &     2023-04-08 &10.52 &   -- &   -- &18.04 &   -- &   --  \\
    Par016 &    152.6594150 &     -4.7508000 &     2023-04-14 &   -- & 0.29 & 0.14 & 1.19 & 0.29 & 0.29  \\
    Par017 &    150.3011090 &      1.8910820 &     2023-04-19 &   -- &   -- & 3.15 &   -- &   -- & 3.15  \\
    Par018 &    159.2102450 &      1.7162860 &     2023-04-20 &   -- & 0.67 &   -- & 1.10 & 1.10 &   --  \\
    Par019 &    130.3650700 &     48.6033460 &     2023-04-20 & 0.10 & 0.10 &   -- & 0.86 & 0.86 &   --  \\
    Par020 &    149.6229230 &      2.2251060 &     2023-04-23 &   -- & 0.13 &   -- & 4.87 & 1.05 &   --  \\
    Par021 &    161.1635170 &     33.9294610 &     2023-04-24 &   -- &   -- & 0.52 & 0.67 & 0.67 & 0.57  \\
    Par022 &    177.2619900 &     22.4367760 &     2023-04-28 &   -- &   -- & 3.15 &   -- &   -- & 3.15  \\
    Par023 &    150.0655290 &      2.5519210 &     2023-05-01 &   -- & 0.13 &   -- & 4.29 & 1.05 &   --  \\
    Par024 &    150.1315240 &      2.4996600 &     2023-05-02 &   -- & 0.13 &   -- & 4.87 & 1.05 &   --  \\
    Par025 &    150.2457730 &      1.9774740 &     2023-05-02 &   -- & 0.13 &   -- & 4.87 & 1.05 &   --  \\
    Par026 &    150.1227170 &      2.4398770 &     2023-05-06 &   -- & 0.13 &   -- & 4.87 & 1.05 &   --  \\
    Par027 &    172.9448010 &    -12.3606500 &     2023-05-09 & 0.29 & 0.29 & 0.29 & 2.58* & 1.15 & 1.15  \\
    Par028 &    150.0874070 &      2.4185310 &     2023-05-20 & 1.79 & 1.43 & 1.43 &21.26* & 7.08* & 5.37*  \\
    Par029 &    150.2522870 &      1.7285570 &     2023-05-22 &   -- & 0.13 &   -- & 4.87 & 1.05 &   --  \\
    Par030 &    216.0875670 &     24.2475070 &     2023-05-25 &   -- & 0.79 &   -- & 0.79 & 0.79 &   --  \\
    Par031 &    187.7406480 &     12.5610560 &     2023-05-31 &   -- &   -- & 0.41 &   -- &   -- & 6.49  \\
    Par032 &    180.4252840 &    -18.8087340 &     2023-06-07 &   -- & 1.00 &   -- & 1.00 & 1.00 &   --  \\
    Par033 &    176.4223950 &     20.3820560 &     2023-06-09 &   -- &   -- & 0.95 &   -- &   -- & 0.95  \\
    Par034 &    190.4031030 &     22.4995390 &     2023-06-21 & 1.24 & 1.24 & 1.24 & 6.20* & 1.24 & 1.24  \\
    Par035 &    181.5579470 &     -8.6302920 &     2023-06-21 &   -- &   -- & 2.98 &   -- &   -- & 2.98  \\
    Par036 &    184.2851990 &      2.7440130 &     2023-06-21 &   -- &   -- & 0.20 &   -- &   -- & 2.03  \\
    Par037 &    180.3574960 &      2.3583210 &     2023-06-23 & 0.48 & 0.48 &   -- & 1.91 & 1.91 &   --  \\
    Par038 &    215.1343710 &     53.0534480 &     2023-06-25 &   -- & 0.13 &   -- & 4.87 & 4.87 &   --  \\
    Par039 &    182.9192700 &     -1.1318990 &     2023-06-29 &   -- &   -- & 0.67 &   -- &   -- & 3.82  \\
    Par040 &    181.3629010 &     -7.5363050 &     2023-07-01 &   -- &   -- & 0.95 & 0.95 & 0.95 & 0.95  \\
    Par041 &    191.4234500 &      3.5618430 &     2023-07-02 &   -- &   -- & 0.43 &   -- &   -- & 3.44  \\
    Par042 &    195.1846320 &     12.5341350 &     2023-07-04 &   -- & 0.57 & 0.57 & 2.58 & 1.00 & 1.00  \\
    Par043 &    213.2233650 &     -3.2187080 &     2023-07-25 &   -- &   -- & 0.14 &   -- &   -- & 0.14  \\
    Par044 &    265.7442370 &     67.0902050 &     2023-09-03 &   -- &   -- & 0.14 & 2.10 & 0.52 & 0.57  \\
    Par045 &    137.7250700 &     -4.4086120 &     2023-11-17 &   -- &   -- & 0.95 &   -- &   -- & 3.01  \\
    Par046 &    149.3468690 &      2.5384070 &     2023-12-03 &   -- &   -- & 0.20 &   -- &   -- & 1.62  \\
    Par047 &    149.4398780 &      2.1723810 &     2023-12-03 &   -- &   -- & 0.20 &   -- &   -- & 1.62  \\
    Par048 &    149.4029130 &      1.6714440 &     2023-12-03 &   -- &   -- & 0.20 &   -- &   -- & 1.62  \\
    Par049 &    149.7024380 &      2.0589470 &     2023-12-03 &   -- &   -- & 0.20 &   -- &   -- & 1.62  \\
    Par050 &    189.1800240 &     62.0654160 &     2023-12-22 &   -- & 2.04 & 2.04 &12.02* & 3.94 & 3.94  \\
    Par051 &    150.5036630 &      2.2633940 &     2024-01-02 & 0.79 &   -- &   -- & 7.09* &   -- &   --  \\
    Par052 &    150.1552210 &      2.0418390 &     2024-01-03 &   -- & 2.04 & 1.36 &12.02* & 3.94 & 3.94  \\
    Par053 &    150.1030700 &      1.7931780 &     2024-01-04 &   -- &   -- & 0.20 &   -- &   -- & 1.62  \\
    Par054 &    214.5759310 &     52.2796830 &     2024-01-15 &   -- &   -- & 4.29 & 4.29 & 4.29 & 4.29  \\
    Par055 &    177.4168450 &     22.1796790 &     2024-01-16 &   -- & 0.79 &   -- & 0.79 & 0.79 &   --  \\
    Par056 &    169.9978930 &      6.5094450 &     2024-01-22 &   -- & 0.79 & 0.79 &   -- & 1.57 & 1.57  \\
    Par057 &    177.4628390 &     22.2952510 &     2024-01-24 &   -- &   -- & 0.67 & 0.91 & 0.91 & 0.67  \\
    Par058 &    177.5042440 &     22.6834370 &     2024-05-16 &   -- &   -- & 4.51 & 4.51 & 4.51 & 4.51  \\
    Par059 &    169.9723860 &      6.7628470 &     2024-05-19 &   -- & 1.43 &   -- & 2.86 & 1.43 &   --  \\
    Par060 &    157.5820920 &      5.5249110 &     2024-05-26 &   -- &   -- & 0.43 &   -- &   -- & 1.15  \\
    Par061 &    159.1929810 &     -2.4619510 &     2024-06-01 & 0.43 & 0.43 &   -- & 2.30* & 2.30* &   --  \\
    Par062 &    188.1922630 &      9.3749900 &     2024-06-20 &   -- &   -- & 0.20 &   -- &   -- & 2.23  \\
    Par063 &    149.9702590 &    -22.8246010 &     2023-12-09 &   -- &   -- & 0.14 &   -- &   -- & 0.14 
\enddata
\tablecomments{The target coordinates represent an average field center, in case of multiple pointings per field. For times marked with an asterisk (*), the total exposure time is split between the G150R and G150C grisms.}
\end{deluxetable*}

\subsection{Emission Line Detection Sensitivity}
In the fields covered by pure parallel visits longer than 3 hours, we usually obtained minimum total integration times of 4800 seconds in two grism filters. The main difference in the longer opportunities was the addition of longer integrations in the F115W filter (by factors of 2 to 3 times). Therefore our sensitivity in the F150W and F200W filters, as shown for some examples in Figure \ref{fig:sensivity}, was
fairly uniform over most of our planned observations: for point sources, the line sensitivity was typically around $3-4\times10^{-18}$ erg/sec/cm$^{2}$ (5$\sigma$). This limiting line flux for H$\alpha$ is equivalent to a Star Formation Rate (SFR) $\sim$0.5 $M_\odot$/yr at $z\sim2$ for a Chabrier IMF \citep{1998ARA&A..36..189K}. Throughout this paper assumes a flat lambda CDM cosmology with $H_0 = 70$ and $\Omega =0.3$.

We confirm the predictions of the exposure time calculator (ETC)\footnote{\href{https://jwst-docs.stsci.edu/jwst-exposure-time-calculator-overview/jwst-etc-calculations-page-overview/jwst-etc-target-acquisition/jwst-etc-niriss-target-acquisition\#gsc.tab=0}{https://jwst-docs.stsci.edu}}; for a spatially extended galaxy (a S\'{e}rsic profile with major and minor axes of $1.0\arcsec \times 0.5\arcsec$, the limiting emission line flux in the central 4 pixels (a $0.13\arcsec \times 0.13\arcsec$ box) was 3 times higher.

These estimates are independently confirmed by published grism spectroscopy in a NIRISS field \citep{Boyett22}.  Although their integrations were somewhat shorter than what we typically obtained, the line flux sensitivities are consistent with our predictions, allowing detections of one or more rest-frame optical emission lines in more than 90\% of their targeted galaxies.

In the shallow fields, we devoted the more limited time to the reddest grism (F200W), since its 1.8--2.3$\micron$ coverage has never been available for any slitless spectroscopy prior to to JWST \citep{2024AAS...24332502C}. The emphasis was on line pairs close in wavelength, viewed at higher redshifts: H$\beta$+[OIII] and H$\alpha$+[SII]$ \lambda$6716,6731\AA.

Contrary to a widely held perception, confusion due to spectral overlap is not generally very serious  during the assembly of this sample of NIRISS slitless spectra. This is because the actual spectra of all objects in a typical high-latitude field cover only a small fraction of the detector area.  The low spectral resolution results in spectra that are usually confined to only 6 x 60 = 360 pixels, which is only 1/10,000 of the total Teledyne H2RG detector area. Therefore, in most PASSAGE observations, only
$\le 15\%$ of the spectrograms suffer from serious overlap with a spectrum of another object.
The second-order spectrograms are longer but about one hundred times fainter than the first order spectrograms. Thus, the only second order spectra that cause significant contamination tend to be from bright galactic stars, which is noted in our output files.  Since we avoided fields of low galactic latitude, this contamination is very minor.

\section{Data Reduction} \label{sec:Data Reduction}
Reduction of grism data is an important challenge not only for JWST,  but also other ongoing and future missions \citep[Euclid, Roman,][]{2023A&A...676A..34E, 2020ApJ...897...98B}. Getting the most out of NIRISS/WFSS observations requires extensive, careful data reduction efforts. For the preliminary reduction of our data, we use the Grism Redshift \&\ Line analysis \citep[\grizli\footnote{\url{https://grizli.readthedocs.io}, version 1.9.5}; ][]{Grizli} software, which is a quantitative and comprehensive modeling and fitting tool for slitless spectroscopic observations. The details of our reduction will be fully described in a future publication, but we include a brief description here.

\grizli\ incorporates the default STScI data reduction pipeline to apply the initial detector-level corrections (performed by \texttt{Detector1} stage of the STScI pipeline\footnote{\url{https://jwst-pipeline.readthedocs.io}}) which includes a variety of steps such as linearity correction, persistence correction, flat-fielding, masking of bad pixels and cosmic rays, dark subtraction, and ramp fitting. \grizli\ further incorporates a custom algorithm for ``snowball" masking\footnote{\url{https://github.com/mpi-astronomy/snowblind}} \citep[\texttt{snowblind}; ][]{Snowblind}. This is followed by \grizli's preprocessing step, which registers the WCS, performs astrometric alignment, subtracts sky background, and drizzles each exposure on a common reference frame to provide fully reduced individual exposures as well as mosaics for both the direct and dispersed images. All of our images are aligned to the Legacy Survey's DR9 \citep{Dey2019} astrometry. The next step is to identify all sources by running ``Source Extraction and Photometry" \citep[\sep\footnote{\url{https://sep.readthedocs.io}}; ][]{SEP,SExtractor} on a detection image typically created by combining all filters where a direct image is available. The absolute flux calibration is performed on individual spectra using the calibration files provided as part of the NGDEEP calibration \citep{Pirzkal24, ravindranath2023niriss}.

In order to model the overlapping spectrograms, which present one of the most challenging aspects of slitless spectroscopy, \grizli\ uses the segmented images of all detected sources (brighter than $m_{AB}$ ranging from 26 to 30, chosen depending on the depth of the direct imaging of the field), and models their associated spectrograms in each individual grism exposure. The pixels associated with an overlapping (contaminating) source are down-weighted when modeling spectral energy distributions (SED) of the primary source of interest and extracting the final grism spectra. Lastly, \grizli\ generates the 2D dispersed spectra by simultaneously modeling each grism exposure, allowing for independent noise and other features such as differences in morphological broadening per exposure. The final 1D spectra generated for each individual object are then modeled using a set of SED templates from \eazy\footnote{\url{https://eazy-py.readthedocs.io}} to assist in redshift identification.

The 0th order contamination is of concern since it can masquerade as an emission line, as discussed below. The 2nd order is only prominent for bright stars and is therefore relatively rare. When modeling the contamination, Grizli flags contributions from both the 0th and 2nd orders, and these pixels are appropriately excluded when fitting the galaxy spectra.

\begin{figure}
\plotone{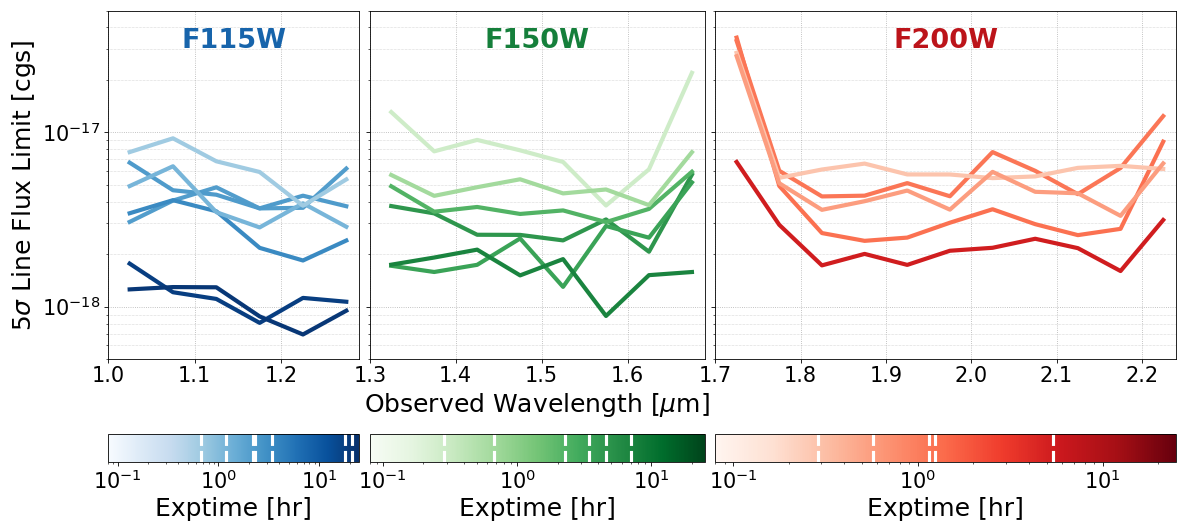}
\caption{Emission Line flux sensitivity in typical PASSAGE observations. The 5-$\sigma$ detection limits, in erg/sec/cm$^{2}$ are shown for point sources in several representative fields, with integration times ranging from our shortest to our longest exposures. }
\label{fig:sensivity}
\end{figure}

For the brightest objects we measure, the systematic uncertainty in the flux scales becomes a significant additional source of error.  Judging from the agreement of independent but overlapping spectra of the same object, our absolute fluxes may sometimes be systematically uncertain by up to 20\%.  The fluxing uncertainty is worse at the blue and red edges of each spectrum.  Grizli does not accurately correct the fluxes where the grism transmission drops below 50\%.  In these small edge regions, we do not attempt to measure emission line fluxes because they are not trustworthy.   Therefore we restrict our flux measurements to these reliably-fluxed spectral windows:  F115W: 1.013 –- 1.283 $\micron$; F150W: 1.330 –- 1.671$\micron$: and F200W: 1.751 –- 2.240$\micron$ (see also Figure \ref{fig:sensivity} and the graphical illustrations of the spectral windows in the one-dimensional plots of Figures 9--14). We warn other users of NIRISS that grism spectroscopy fluxes beyond these wavelength regions should not be used. 

Experience with NIRISS grism spectra reduced by the \grizli\ software package showed that it picks up a significant number of artifacts (e.g., emission lines from other nearby sources, diffraction spikes, other image artifacts, modeling issues with the continuum, overlapping spectra or zeroth orders, etc.). These were incorrectly identified as real sources, which must be removed by (our) human intervention. Another problem is that when spectra of two objects partially overlap, \grizli's attempt to automatically remove the contamination often leads to over-subtraction. This usually causes parts of the overlapping spectra to (artificially) drop below zero flux. Efforts to replace some of this labor intensive work with computer algorithms are ongoing.

\begin{figure}
  \centering
  \includegraphics[angle=90, width=0.8\linewidth]{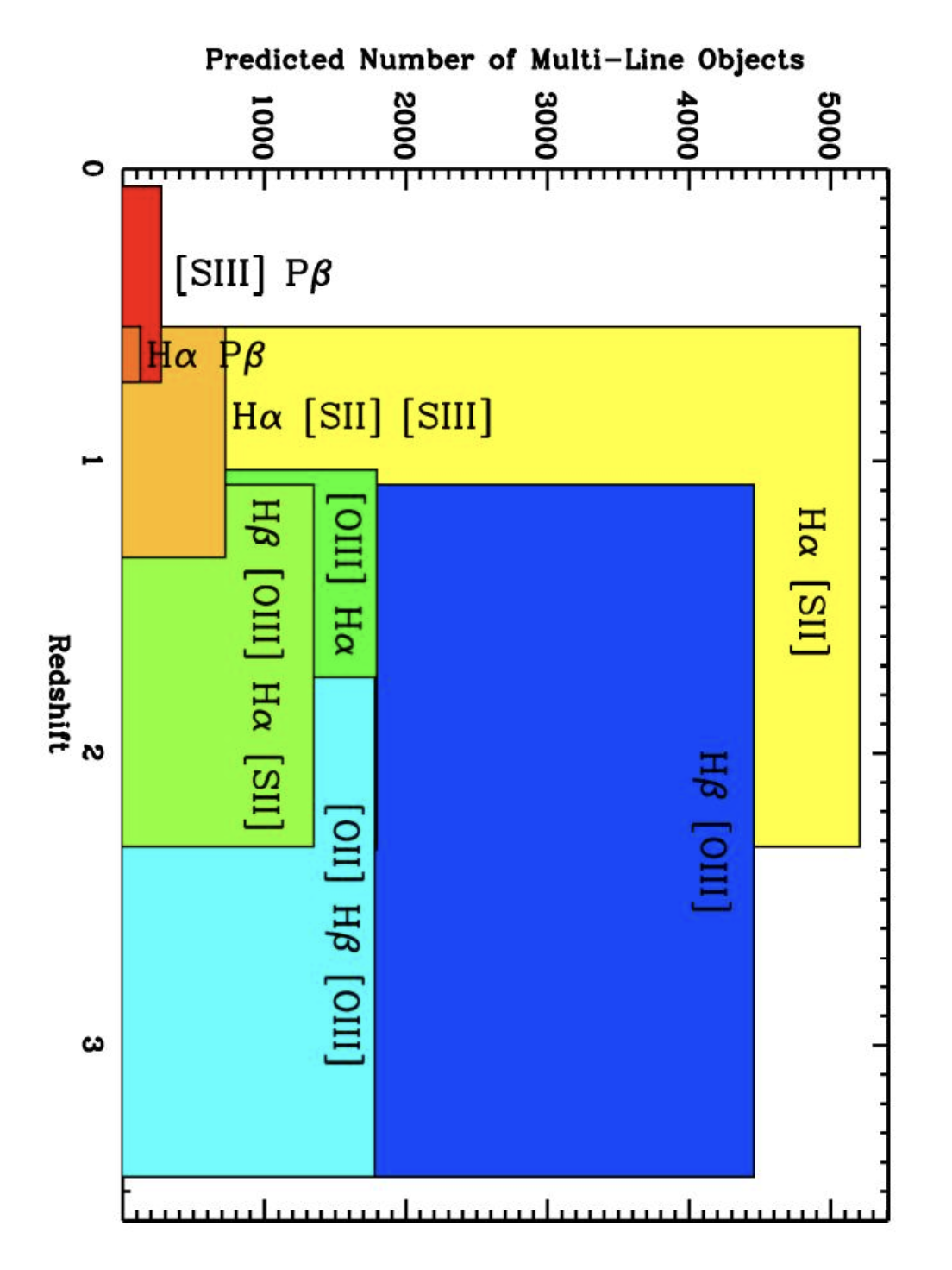} 
  \caption{ Total predicted multiple emission line NIRISS galaxies for the 63 fields from PASSAGE, Cycle 1. All lines are detected at 5$\sigma$. The height of each box is the expected number of emitters and the width is the span of redshift. Each box is labeled with the set of multiple emission lines it represents. Not all possible line combinations are shown. There will also be many more multiple emission line sources where only a single line will be detected at 5$\sigma$ significance, but other lines will be robustly measurable at lower significance. The primary emission lines we use for redshift determinations, H$\alpha$/[SII] and/or [OIII]/H$\beta$, are predicted to be available in almost ten thousand galaxies.  This estimate is confirmed by our detailed inspection of the first PASSAGE fields.}

  \label{fig:multiline_counts}
\end{figure}

\section{PASSAGE Science Examples With Faint Emission-Line Galaxies} \label{sec:faint Emission Line}

The resulting PASSAGE spectra are of great value for investigating many questions of strong current interest in research on galaxy evolution. Our predicted number of emission line-determined redshifts for the entire PASSAGE survey is shown in Figure \ref{fig:multiline_counts}. We now provide a few examples of relevant emission-line galaxy properties at intermediate redshifts that we are measuring with PASSAGE data.

\subsection{Star Formation Main Sequence}\label{SFMS}

According to equilibrium growth models \citep{Lilly2013, Tacconi2020}, galaxy evolution is controlled by cosmic accretion, merging, galactic gas depletion and star formation, and gas recycling into the circumgalactic medium.
Models of these mechanisms aim to reproduce the observed
scaling relations between galaxy stellar mass, star formation rate (SFR), metallicity, gas content, size, and structure \citep[e.g.,][]{Noeske2007, Maiolino2019, Saintonge2011, vanderWel2014, Nedkova2021MNRAS},
along with their evolution over cosmic time. Observed relations appear to show large scatter, whose origin is debated. This can be modeled by short-term stochasticity, long-term differences in growth histories or gas fractions, or different scenarios for feedback from AGN and star formation \citep{Lilly2013, Ma2016, DeRossi2017, Torrey2018, Torrey2019, Dave2017, Dave2019}. At low masses, it is unknown whether these relations change or break down completely \citep{Atek2014c, Alavi2016, Amorin2017}.

Unlike massive galaxies, low-mass ones often have high specific SFRs (normalized by stellar mass), with extremely strong high-ionization emission lines, elevated ionizing power, and perhaps low dust content. To test and constrain the stochastic star formation histories of hydrodynamical simulations \citep{Shen2014, Dominguez2015, Sparre2017, 2024MNRAS.tmp.2365B}, 
we are deriving SFRs from the Balmer emission lines, which reflect the star formation activity on a few Myr timscale. The SFR is derived from the H$\alpha$ emission line, following the calibration of \citet{kennicutt12}. Where available, the Balmer line ratios(e.g., H$\alpha$/H$\beta$ allow us to estimate the effects of dust redenning. In our representative medium field, we measured the Balmer decrement in a sample of 122 galaxies. We derived a median value of E(B-V) ranging from $0.01$ to $0.02$, using the \citet{cardelli89} attenuation law or the SMC extinction law \citep{gordon03}, respectively.  

\begin{figure}[htbp]
  \centering
  \includegraphics[width=0.5\linewidth]{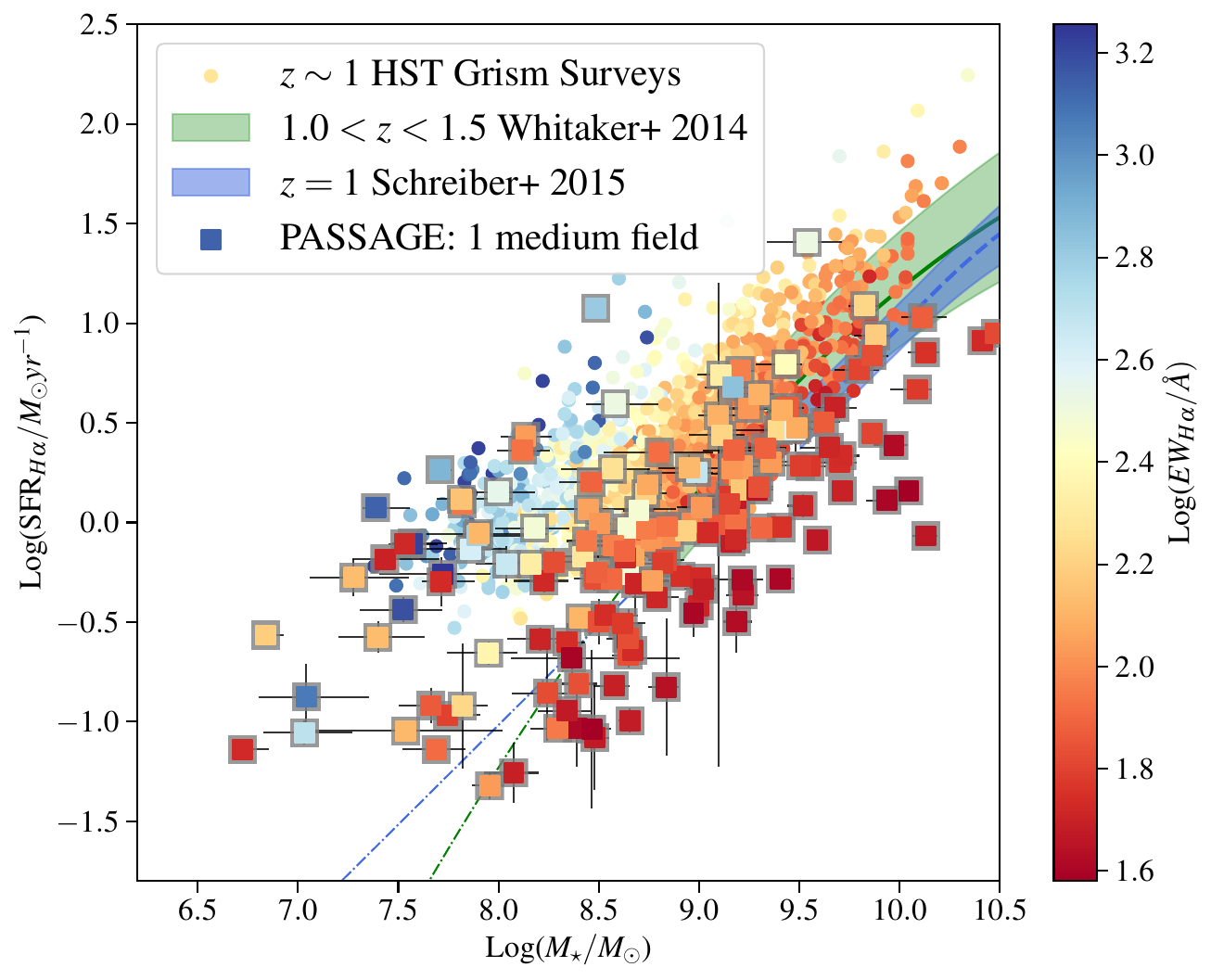} 
  \caption{Massive galaxies follow the main sequence SFR-$M_{\star}$ relation (SFMS) at $z\approx1$, but low-mass galaxies scatter upward to very high H$\alpha$ equivalent widths.  Flux incompleteness at low-masses ($M_{\star} \sim 10^{8} M_{\odot}$) flattens the slope of the relation. In addition to expanding the low-mass sample size by an order of magnitude, PASSAGE extends the mass regimes down to $10^{7}M_\odot$ at $z\sim2$.  The colored squares show detections from a \textit{single} publicly released NIRISS/WFSS field of medium depth.  Stellar masses were estimated from SED fitting, using PASSAGE and COSMOS photometry. SFRs were obtained from H$\alpha$ luminosity, using the \citet{kennicutt12} calibration. Error bars show the statistical uncertainties.}

  \label{fig:sfr_mass}
\end{figure}

Figure~\ref{fig:sfr_mass} presents the star-formation main sequence (SFMS) of galaxies at $z\sim1-2$ in a representative medium-depth field. Stellar masses were estimated via SED fitting, using a combination of our PASSAGE broadband photometry and COSMOS-Web imaging, which covers this field (see Section \ref{sec:reddening}). Out of the 63 PASSAGE fields, 18 have overlapping coverage with COSMOS-Web (Figure \ref{fig:cosmosoverlap}).

Photometric fluxes were obtained within 1\arcsec\ apertures across the following bands: HST/ACS F814W; JWST/NIRISS F115W, F150W, F200W; JWST/NIRCam F115W, F150W, F277W, F444W; and JWST/MIRI F770W. 
We used the \textsc{Bagpipes} software\citep{Carnall2018} to perform the SED (spectral energy distribution) fitting.
We adopted a non-parametric star-formation history using the continuity prior of \cite{2019ApJ...876....3L}, with five logarithmically-spaced age bins (fixing the most recent to cover 0--30\,Myr).
The metallicity was allowed to vary between 0.01 and 2.5\,$\mathrm{Z}_{\sun}$, and the logarithm of the ionization parameter between $-3.5$ and $-2$.
Nebular emission was included for populations as old as 20\,Myr, and dust attenuation was parameterized following \cite{cardelli89}, with $R_V=3.1$, and $A_V$ varying between zero and two magnitudes.
Star formation rates, SFR$_{\rm{H}\alpha}$, were computed using the \citet{kennicutt12} calibration and are corrected for the small dust attenuation inferred from their Balmer decrements.

Our results in figure \ref{fig:sfr_mass} show that firstly, at the higher redshifts of many PASSAGE galaxies ($z > 1$), they tend to have higher specific star formation rates, resulting in higher H$\alpha$ equivalent widths. Secondly the observed SFMS increases towards lower masses and higher H$\alpha$ equivalent width, therefore PASSAGE is particularly sensitive to very low-mass galaxies which are undergoing intense bursts of star formation.

\begin{figure}[htbp]
  \centering
  \includegraphics[width=0.6\linewidth]{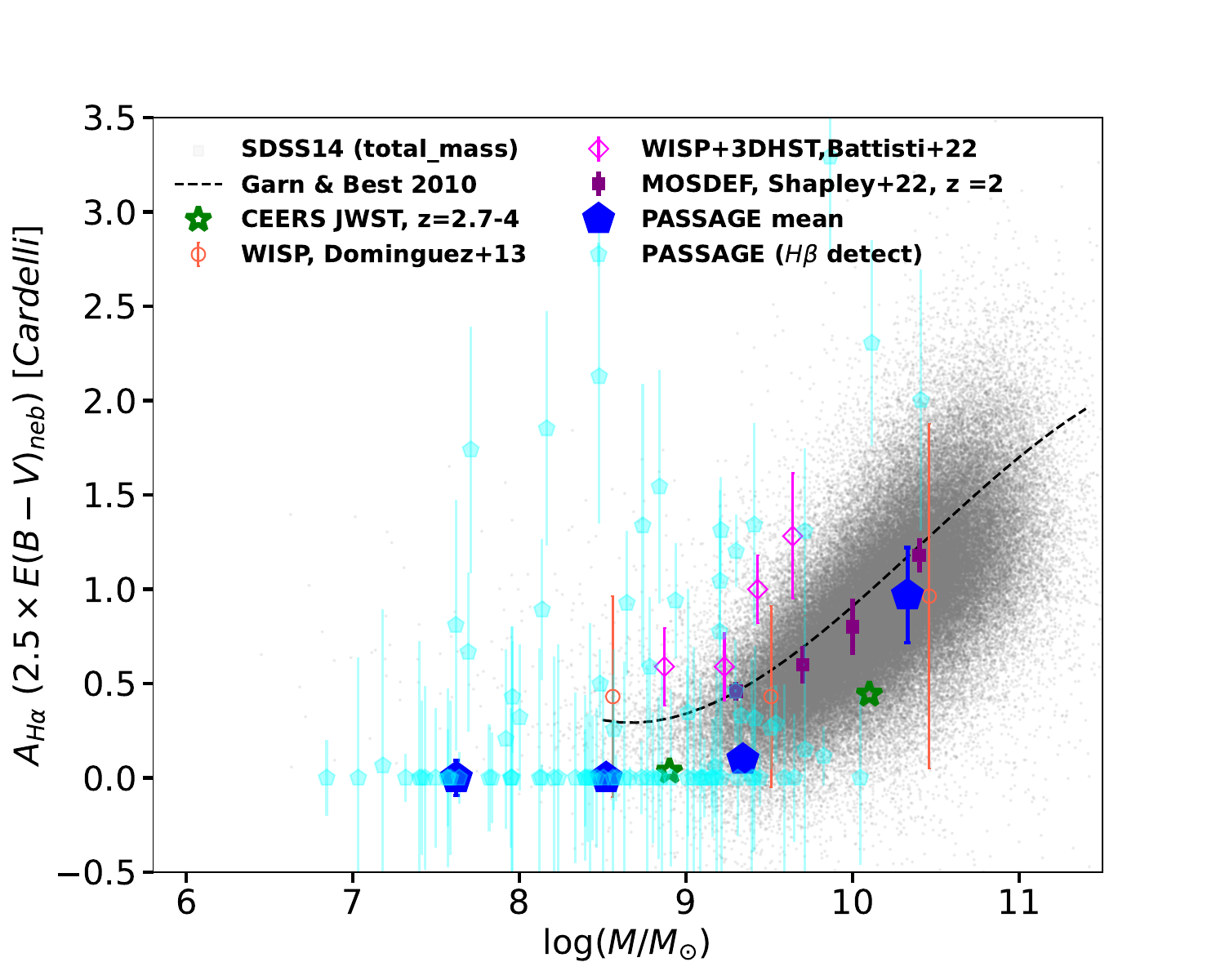}
  \caption{Correlation of gas extinction (i.e., from Balmer decrement) with stellar mass. Error bars show the statistical uncertainties. Most PASSAGE dwarf galaxies have small or negligible extinction, but a significant minority are substantially dusty. On average, our new low-mass galaxies continue the trends found in somewhat more massive galaxies by the WISP survey and substantially more massive galaxies in MOSDEF \citep{Shapley2022}.  Studies such as \citet{Battisti22, Lorenz2024, Dominguez2013} highlight an offset in the dust attenuation–mass relationship depending on whether the fiber-region stellar mass or the total stellar mass is considered for SDSS galaxies. For the SDSS galaxies, we have used the total stellar mass.}
  \label{fig:balmer decrement}
\end{figure}

\begin{figure}[htbp]
    \centering
    \includegraphics[width=1\textwidth]{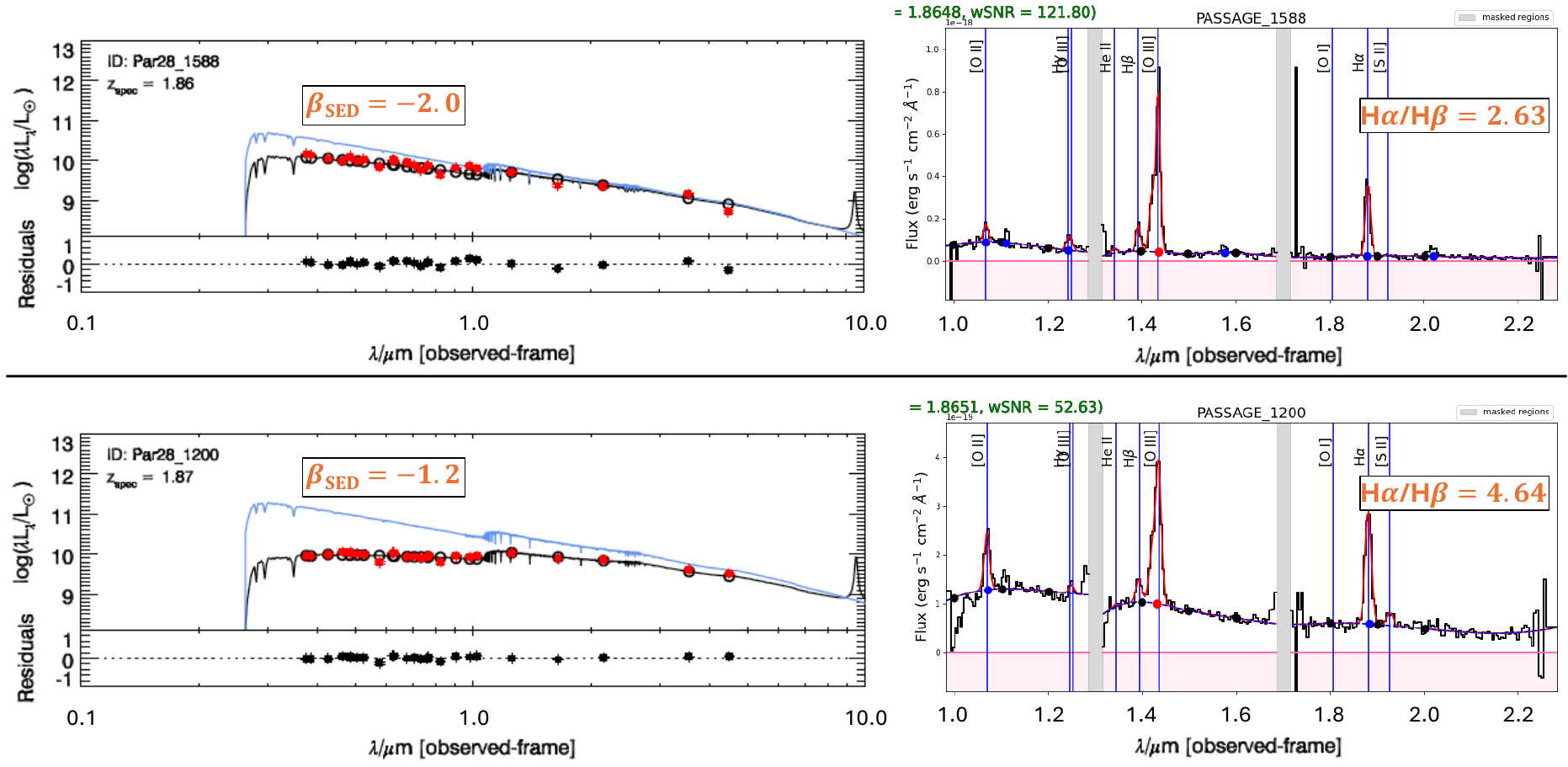}
    \caption{Comparison of the reddening on the stellar continuum UV slope based on an SED fit, $\beta_\mathrm{SED}$ ($F_\lambda\propto\lambda_\mathrm{rest}^\beta$; $1258\text{\AA}\leq \lambda_\mathrm{rest}\leq2580$\AA),relative to the Balmer decrement, H$\alpha$/H$\beta$ for two galaxies at $z=1.9$ in Par28. Left panels show an SED fit (black line) to COSMOS2020 photometry \citep[red symbols;][]{weaver22} using \texttt{MAGPHYS} \citep[][mainly to demonstrate the wide rest-frame SED coverage for PASSAGE fields in COSMOS]{daCunha08, battisti20}. The blue line is the prediction for the unattenuated stellar continuum. Right panels show the PASSAGE grism spectra. The upper galaxy has a lower Balmer decrement relative to the lower galaxy, indicating that the lower galaxy is experiencing more dust attenuation. This is reflected in the values of the UV slopes, with the upper-left panel having a bluer UV slope relative to the lower-left panel. By comparing SEDs for a large number of galaxies with Balmer decrements, we can derive the average dust attenuation curves and their evolution with galaxy properties.}
  \label{fig:Par28_Balmer}
\end{figure}

\begin{figure}
    \centering
    \fig{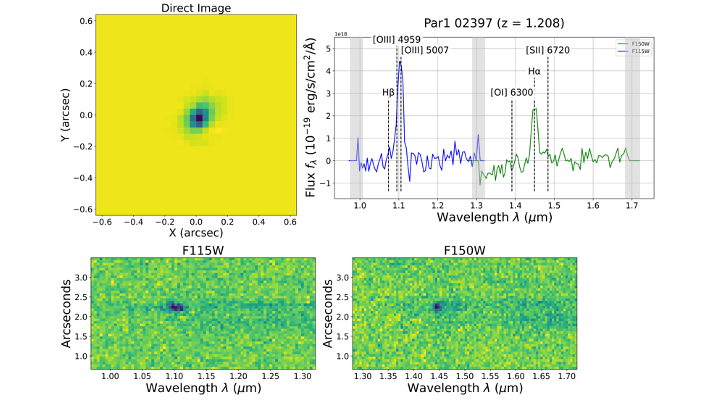}{0.85\textwidth}{}
    \caption{Example of an extreme emission line galaxy (EELG) from the Par~1 field. The upper left panel shows the broadband direct image, 1.28\arcsec\ on a side (32~$\times$~32 pixels at 0.04\arcsec/pixel), drizzled with \texttt{grizli} and oriented with North up. The lower rectangular panel shows the two-dimensional spectrogram at the native 0.066\arcsec/pixel scale, and the upper right displays the extracted one-dimensional spectrum. The flux scale is unreliable near the blue/red ends of each filter, shown by the full vertical grey rectangles.}
    \label{fig:par1_2397}
\end{figure}

\begin{figure}
    \centering
    \fig{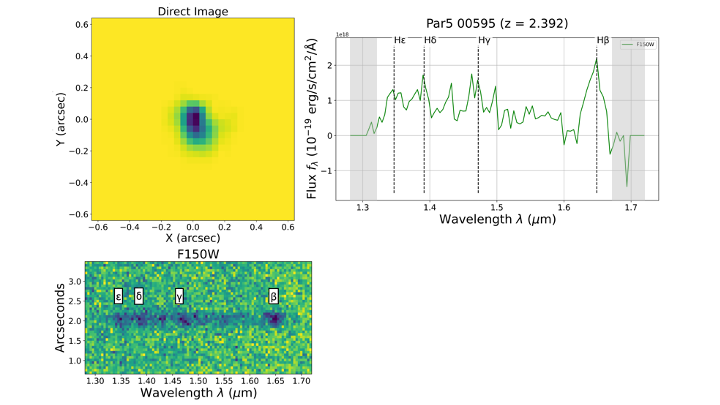}{0.85\textwidth}{}
    \caption{Another example of an extreme emission line galaxy, displayed in the same layout as Figure~\ref{fig:par1_2397}. This compact galaxy from the Par~5 field shows a prominent Balmer series, with emission visible up to H$\epsilon$.}
    \label{fig:par5_595}
\end{figure}

\begin{figure}
    \centering
    \fig{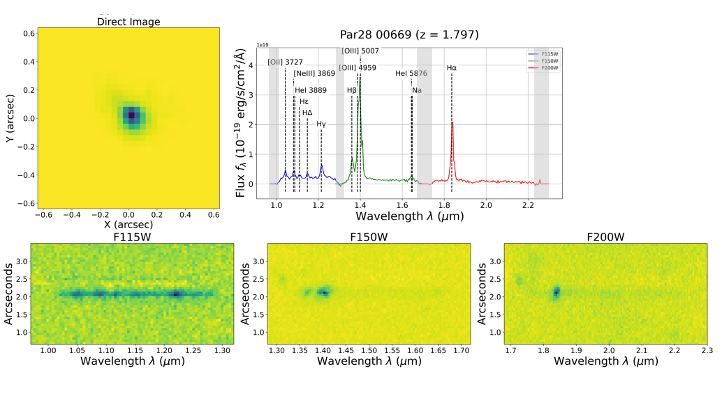}{0.85\textwidth}{}
    \caption{A line-rich EELG from the Par~28 field, shown in the same format as Figure~\ref{fig:par1_2397}. The object exhibits strong [O\,\textsc{iii}], [O\,\textsc{ii}], and Balmer emission from H$\gamma$ through H$\alpha$, visible across all three grism filters.}
    \label{fig:par28_669}
\end{figure}

\begin{figure}
    \centering
    \fig{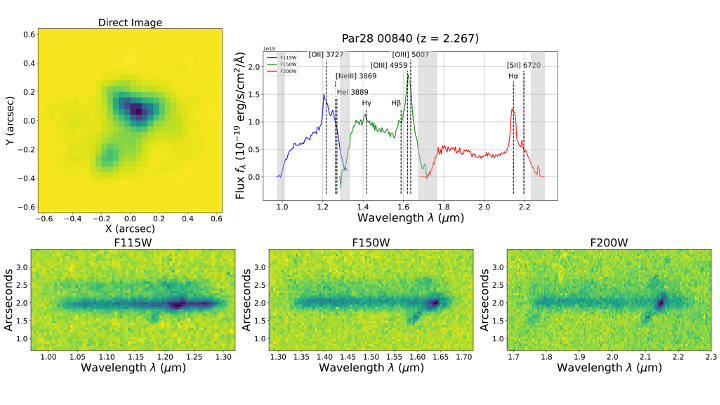}{0.85\textwidth}{}
    \caption{A clumpy or interacting EELG from Par~28, displayed in the same layout as Figure~\ref{fig:par1_2397}. The disturbed morphology and extended emission in the 2D spectrum suggest merging components.}
    \label{fig:par28_840}
\end{figure}

\begin{figure}
    \centering
    \fig{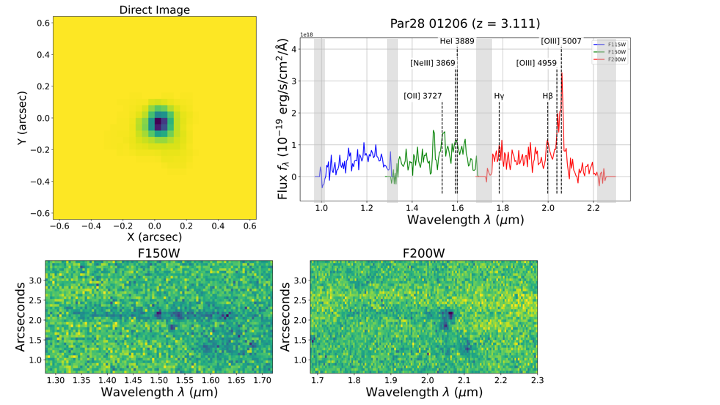}{0.85\textwidth}{}
    \caption{A compact galaxy at $z \approx 3$ from Par~28, shown in the same format as Figure~\ref{fig:par1_2397}. It displays moderate emission from [O\,\textsc{ii}], [O\,\textsc{iii}], and H$\beta$, offering useful spectral diagnostics at intermediate redshift.}
    \label{fig:par28_1206}
\end{figure}

\begin{figure}
    \centering
    \fig{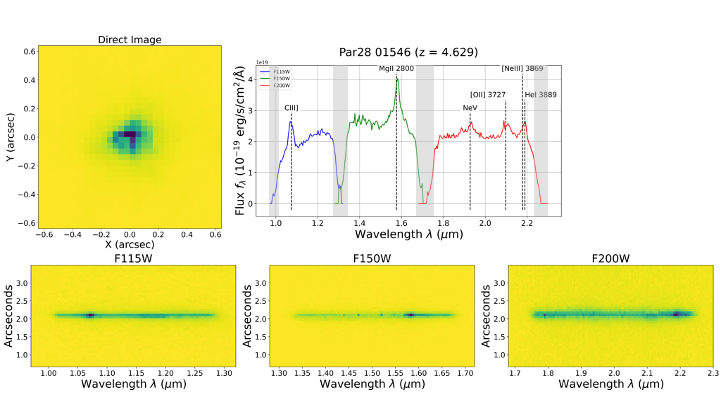}{0.85\textwidth}{}
    \caption{A high-redshift ($z \approx 4.6$) compact galaxy from Par~28, shown in the same format as Figure~\ref{fig:par1_2397}. Strong broad rest-frame UV features include C\,\textsc{iii}]~$\lambda1909$ and Mg\,\textsc{ii}~$\lambda2800$ indicate this is a quasar.}
    \label{fig:par28_1546}
\end{figure}

\subsection{Dust Reddening}
\label{sec:reddening}

Dust attenuation traced by the Balmer decrement, H$\alpha$/H$\beta$, has been shown to correlate with various galaxy properties such as stellar mass, gas-phase metallicity, SFR \citep{Garn2010, Zahid2014, Shapley2022, Battisti22} and more recently with velocity dispersion \citep{maheson2024}. Among these properties, the primary dependency has been found to be with stellar mass \citep{maheson2024}. Recently, \citet{Shapley2023} used JWST/NIRSpec data from the CEERS program to study the dust attenuation-stellar mass relation at higher redshifts $z=3-6$ \citep{2024arXiv240717551R}. However, the positive correlation between dust attenuation and stellar mass has mostly been studied in galaxies with $M_\ast> 10^{9} M_\odot$. Existing studies of dust attenuation in lower mass galaxies are primarily based on stacks of low-resolution, low-SNR spectra (e.g., \citealt{Dominguez2013, Battisti22}).

PASSAGE is obtaining hundreds (and ultimately a few thousand) of individual ``gold standard'' gas extinction measurements using the H$\alpha$/H$\beta$ ratio at $1.1\lesssim z\lesssim 2.4$, down to stellar masses of a few $10^{7}\ \text{M}_{\odot}$, $10\times$ less massive than current measurements \citep{Reddy2015, Dominguez2013, Battisti22}, as seen in Figure \ref{fig:balmer decrement}. To account statistically for blending of [NII] with H$\alpha$, we can rely on established calibrations for the [NII]/H$\alpha$ ratio as functions of stellar mass and redshift \citep{faisst18}, which are calibrated up to $z\sim2.5$, and have been used previously for the purpose of measuring Balmer decrements from low spectral resolution grism data \citep[e.g.,][]{Battisti22}. We note that the [NII] correction is not applied to the PASSAGE measurements in Figure \ref{fig:balmer decrement} as most galaxies have low stellar masses, where [NII] contribution is $<10\%$ of H$\alpha$ flux \citep{faisst18}. PASSAGE detects large
numbers of extreme emission line galaxies (EW [OIII] $> 300 \text{\AA}$). Their starlight continuum is extremely faint, suggesting stellar masses as low as log $M_*/M_\odot \sim 7$ that have been missed by most previous surveys. 
Figure \ref{fig:balmer decrement} shows the correlation of gas extinction (from the Balmer decrement) with stellar mass. Most PASSAGE dwarf galaxies have negligible extinction, but a notable minority are substantially dusty. We note that the individual PASSAGE galaxies shown here are selected to have at least a $3\sigma$ detection in H$\beta$, which can bias the sample toward lower attenuation. This is evident in the figure, where many galaxies show negligible or zero attenuation. The Balmer line ratios used here are corrected for Balmer absorption lines from the atmosphere of stars, mainly A-type. We used an average stellar absorption corrections of $1.4\%$ and $11.8\%$ for H$\alpha$ and H$\beta$, respectively \citep[Alavi et al. 2025 (submitted)][]{Reddy2015}.

For fields with ancillary photometric data, we can combine Balmer decrement with SED model fits to this photometry, to derive average dust attenuation curves (\citealt{Reddy2015, Battisti22}). Such data are already available for 18 PASSAGE fields (29\%) located within the COSMOS footprint, and more will be available in the future from follow-up programs with $HST$ (7 non-COSMOS fields; PI:Mehta, \#17461), VLT/FORS2 (9 non-COSMOS fields; PI:Battisti, 114.27EF.001, 115.27WC.001/115.27WC.003), and Keck/LRIS (2 non-COSMOS fields with grism data in all three filters; PI:Nedkova, 64/2025A\_N067). 
This is illustrated in
Figure \ref{fig:Par28_Balmer}, which contrasts examples of an emission-line galaxy (top) with little if any extinction in either its starlight or gas, with another galaxy (bottom) having substantial continuum and emission-line reddenings. We will further constrain the optical shape of the dust extinction curve using higher-order Balmer lines (via stacking; eg., \citealt{Reddy2020}).

\begin{figure}[htbp]
  \centering
  \includegraphics[width=0.8\linewidth]{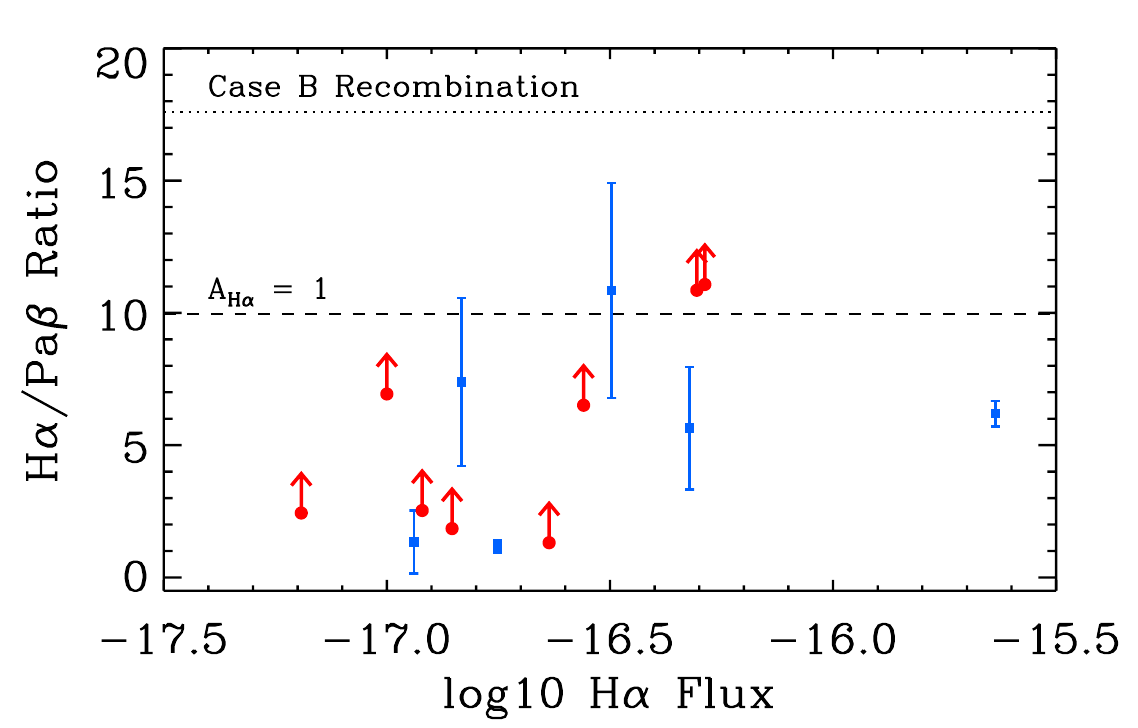}
  \vspace{-5pt} 
  \caption{PASSAGE measurements of the reddening-sensitive Balmer to Paschen emission line ratio in a small sample of galaxies at $z \le 0.7$. These are plotted as a function of H$\alpha$ line flux, in cgs units. The dotted horizontal line shows the predicted intrinsic ration of H$\alpha$/Pa$\beta$ for Case B recombination with no extinction A$_{\text{H}\alpha}$ = 0.  Some detections and upper limits are consistent with the H$\alpha$ to Pa$\beta$ ratio of 10, which is expected based on the typically assumed one magnitude of extinction at the wavelength of H$\alpha$ (shown by the dashed line). However several galaxies show significantly weaker H$\alpha$, indicative of higher reddening.}
  \vspace{-10pt} 
  \label{fig:Pa/Hb}
\end{figure}

At lower redshifts ($z \le 0.7$), our longest wavelength spectral coverage includes the Pa$\beta$ recombination line. We show the ratio of H$\alpha$ to Pa$\beta$ for individual galaxies in Figure \ref{fig:Pa/Hb}.  Not surprisingly, no galaxies have observed ratios comparable to the intrinsic ratio predicted by pure Case B recombination (top dotted line). Instead, as expected, many galaxies have significantly weaker H$\alpha$, which is in most cases consistent with the standard expectation of 1 magnitude of absorption at H$\alpha$ \citep{Kennicutt1983}. Interestingly, a significant fraction (4 out of 8) of the galaxies have H$\alpha$ even considerably weaker, indicative of 2 or more magnitudes of dust extinction.

\begin{figure}[htbp]
  \centering
  \includegraphics[width=0.6\linewidth]{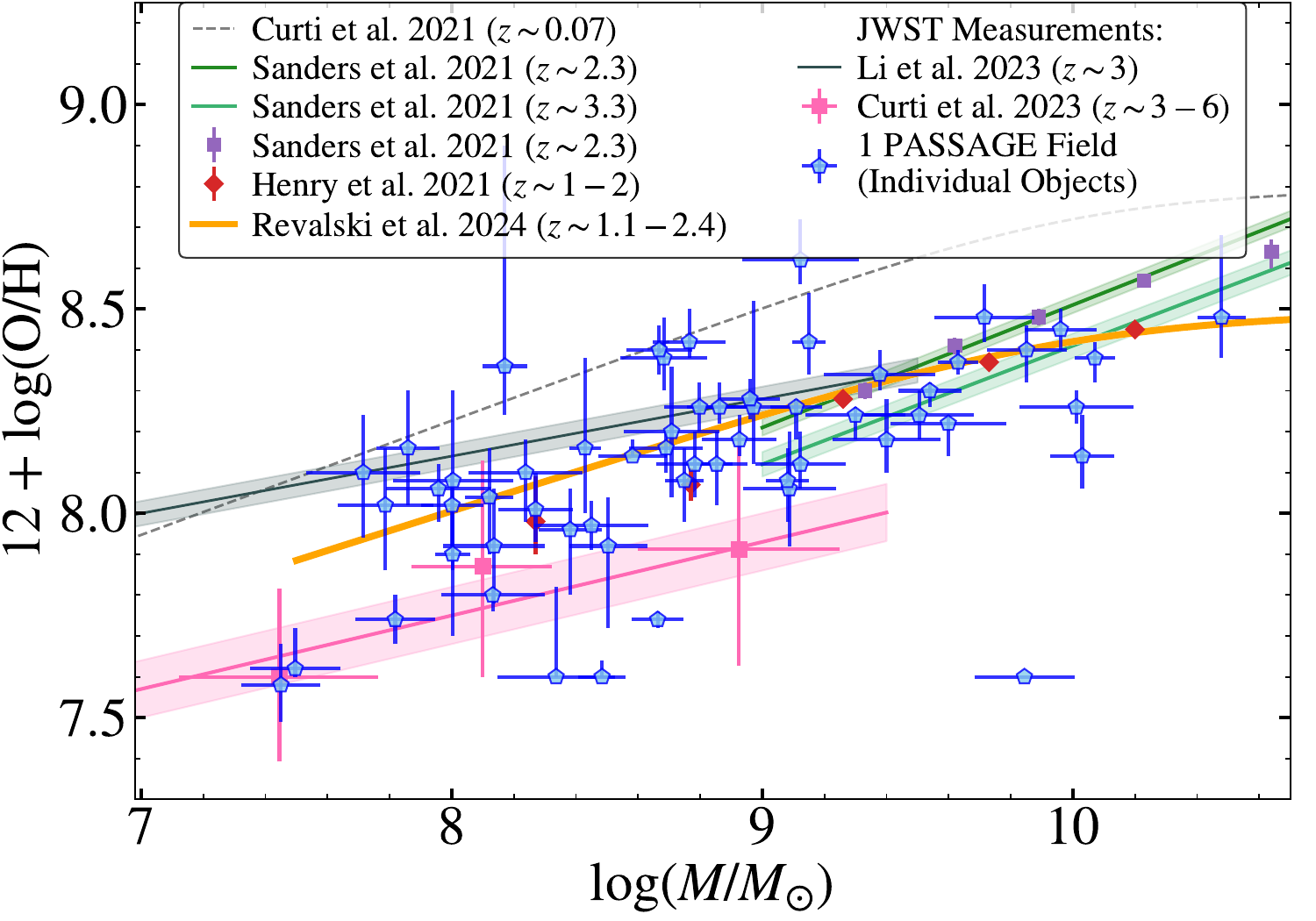} 
  \caption{Gas-phase mass-metallicity relation of individual galaxies from one PASSAGE field compared to stacked results from the literature \protect\citep{Curti2017MNRAS, Sanders2021, Henry2021, Li2022_MZR, Curti2023MNRAS}. A single PASSAGE field already extends the relation to low stellar mass objects at $1.7 \leq z \leq 3.4$.}
  \label{fig:MZR}
\end{figure}

\subsection{Extreme Emission Line Galaxies}
The slitless spectroscopy of PASSAGE is able to detect emission lines independently of the underlying continuum brightness.
Thus in contrast to typical magnitude-limited spectral surveys, PASSAGE selects an emission-line flux-limited sample of galaxies, rather than a continuum-flux limited sample.
PASSAGE can measure galaxies with emission lines of extremely high-equivalent widths, even when the continuum is very faint. This results in a large new sample of so called ``extreme emission line galaxies", which have line equivalent widths exceeding 200 \text{\AA} to more than 1000 \text{\AA} \citep{2014ApJ...789...96A, 2011ApJ...743..121A}. Such extreme but faint galaxies tend to be excluded from traditional spectroscopic surveys \citep{Chen2024}.

Figures \ref{fig:par1_2397}, \ref{fig:par5_595}, 
\ref{fig:par28_669},
\ref{fig:par28_840}, and
\ref{fig:par28_1206}
show examples of PASSAGE extreme emission line galaxies (EELGs). They have H$\alpha$ emission lines with rest-frame equivalent widths of a few hundred Angstroms or more, indicative of extremely high specific star formation rates. This corresponds to extremely high specific star formation rates larger than $10^{-8}$ $yr^{-1}$. These imply that much of the star formation in these galaxies happened within a hundred million years of the epoch at which we observe them \citep{Atek2014c}. The low spectral resolution of the NIRISS grism blends in the [NII]6584/6548\AA~doublet with H$\alpha$. However we can remove the [NII] contamination with statistical corrections, which indicate that it is a small contributor in typical EELGs. The corresponding timescale to double their stellar masses may be only $\sim10^8$ years \citep{Atek2014c, Boyett22}.  Furthermore, their intense bursts of star formation are often concentrated into extremely small volumes \citep{2011ApJ...743..121A}. It is not always easy to classify these highly disturbed dynamical systems into individual galaxies, or assembling galaxy fragments.

PASSAGE finds as many as a hundred such extreme H$\alpha$-emitters per field, out to redshifts of $z \sim 2.5$. For example Figure \ref{fig:sfr_mass} shows that galaxies with EW of H$\alpha>200\text{\AA}$, the light blue and dark blue points \textit{dominate} the PASSAGE sample of spectra in this single field.
Figure \ref{fig:sfr_mass} also illustrates that larger numbers of extreme emission line galaxies can be inferred from the detection in PASSAGE spectra of [OIII] line emission of very high equivalent width out to $z \sim 3.5$. And as illustrated by the example spectra in Figure \ref{fig:par28_1546}, the strong UV-rest frame emission lines of quasars can be seen in PASSAGE at higher redshifts, out to $z > 5$.

\subsection{Gas-Phase Mass-Metallicity Relations}
The mass-metallicity relation (MZR) is the correlation between galaxies’ stellar masses and their gas-phase metallicities. Constraining this relation over large ranges of stellar mass, and its evolution with redshift, provides crucial constraints on models of gas accretion, star formation, metal enrichment, and outflows, which are responsible for regulating galaxy growth. 
PASSAGE includes many low-mass galaxies, reaching largely
unobserved ranges of stellar mass, gas fraction, and redshift
(Figure \ref{fig:MZR}, \citealt{Sanders2018, Henry2021, Sanders2021, wangMassMetallicityRelation2022, Curti2023MNRAS, Revalski2024ApJ}).
At low stellar masses and high redshifts, outflows may preferentially remove metals \citep{Henry2013}, thus making the low mass MZR slope a key diagnostic of the role of gas recycling at different epochs, quantifying the role of feedback and of the stochastic processes in regulating metallicity. PASSAGE, due to its large sample size, is also measuring the scatter in the MZR for low-mass galaxies \citep[see e.g.,][]{Torrey2019}.

Prior to JWST, these regimes had not been constrained because of the low number and quality of spectra below 10$^{9} M_\odot$, which required stacking.  
More recent results with NIRISS \citep{Li2022_MZR,heEarlyResultsGLASSJWST2024} have shown the feasibility of metallicity measurements at lower masses, albeit with uncertain values and scatter.
PASSAGE will address this by measuring gas metallicities of many individual low-mass galaxies---without requiring stacking---up to $z$=3.5.

For the several thousand galaxies in the redshift range $1.7\lesssim z \lesssim 3.5$ for which the multiwavelength coverage of the PASSAGE grism filters allows simultaneous detection of [OII], [OIII], and H$\beta$ lines, we are able to determine the MZR using strong line diagnostics such as the $R_{23}$ index \citep[e.g.,][]{Pagel1979MNRAS, Tremonti2004ApJ}. In Figure~\ref{fig:MZR}, we show the position of 58 individual galaxies from a single deep PASSAGE field on the stellar mass--metallicity plane. These objects are selected to have S/N$\geq 10$ \OIII detections. Metallicities are obtained using the \cite{Curti2017MNRAS} strong line calibrations, which have been validated out to at least $z\sim2.5$ \citep{Revalski2024ApJ} and their stellar masses are derived as described in \S\ref{SFMS}.

Due to the long-lasting interplay between the processes that drive the baryon cycle, the metallicity of galaxies has been also observed to anti-correlate with star formation rate at fixed stellar mass. This relation is commonly referred to as the ``Fundamental Metallicity Relation" (FMR) \citep[FMR; e.g.][]{Ellison2008ApJ, Mannucci2010, Salim2015ApJ, Telford2016ApJ, Baker2023MNRAS.519.1149B}. PASSAGE is now establishing if high redshift galaxies follow the FMR, as at $z$$\sim$0 \citep{Mannucci2010}.

\subsection{Absorption Line Measurements.} 
Stellar age and abundance measurements from absorption-line spectroscopy generally require rest-frame coverage from 3700\AA\ to 5300\AA. PASSAGE obtains this at $1.6<z<2$ for all medium and long visits, and at $2.0<z<3.3$ for long visits which include the F200W spectral band. In each field
we typically find around a few spectra of $z=$1.6--2.5 quiescent galaxies which are bright enough
( $m_{F115W}\lesssim24.5$)
to have very high SNR$_{F115W}$ (ten or larger) per spectral pixel.
In these cases, although emission lines are either weak or undetected, we can measure diagnostic {\it stellar absorption} features.
If we restrict consideration to the strongest stellar absorptions, the Balmer and Ca II HK breaks--these can be measured even when the continuum is one to two magnitudes fainter, giving us an average of $\sim 5$ such spectra per field.

\begin{figure}[htbp]
    \centering
    \includegraphics[width=\textwidth]{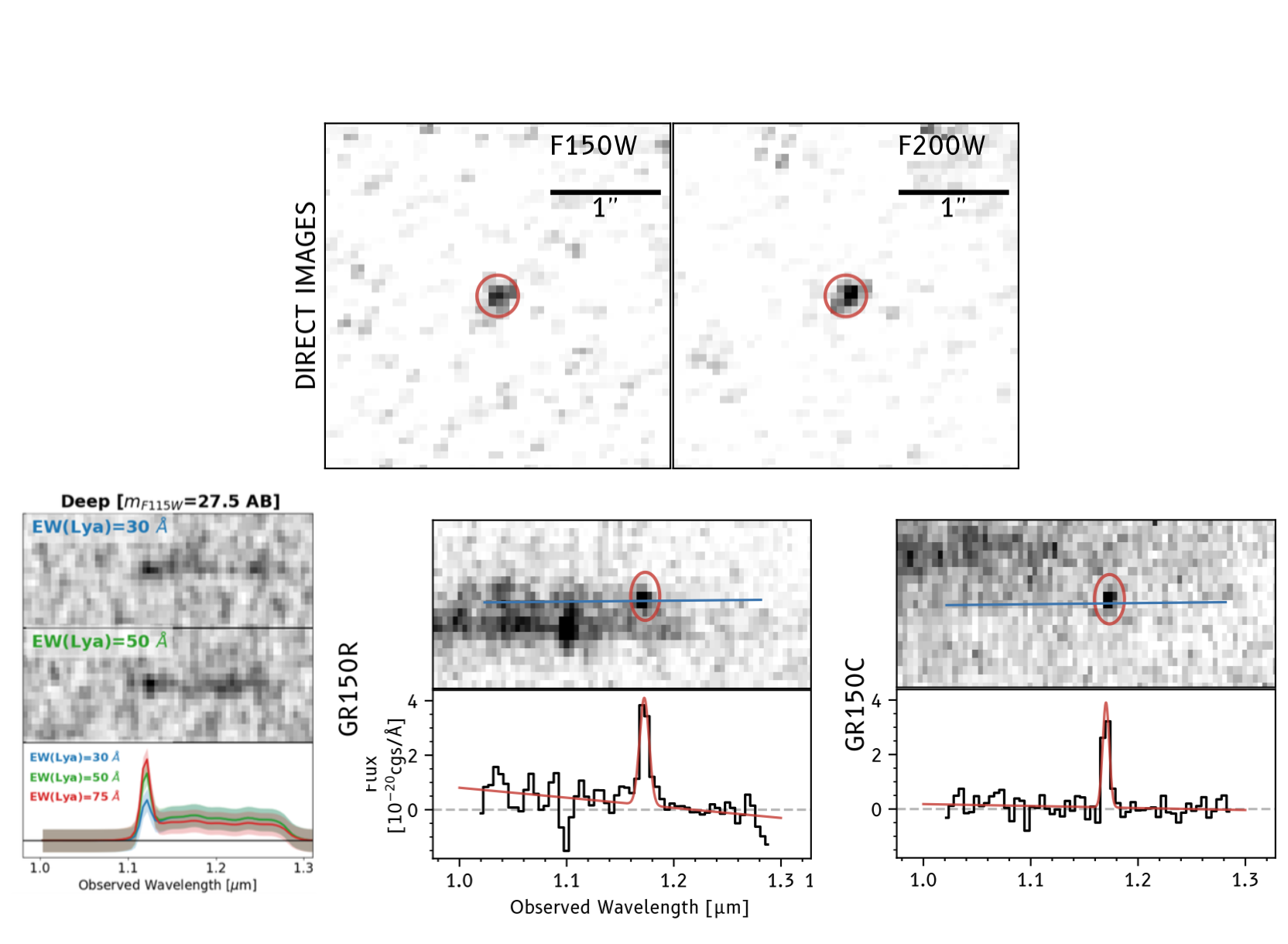}
    \caption{
    PASSAGE is able to measure Ly$\alpha$ emission (LAE) with rest-frame EW $> 50 \text{\AA}$ in spectroscopically confirmed $z > 8$ galaxies. The lower left-side panel shows simulated $z = 8.2$ LAE spectra with $m_{\text{F115W}} =  27.5$ (deep) with rest-frame EW(Ly$\alpha$) of 30 \text{\AA} (top row) and 50 \text{\AA} (middle row). These are predicted to have clear Ly$\alpha$ detections. The bottom sub-panel shows the simulated 1D spectra along with their noise for a range of EWs.  
    These simulations are then compared with an actual PASSAGE detection of Ly$\alpha$ line emission.  The lower right-side panels show the 2-d and 1-d F115W spectra for the object (visible in the direct images above), for both grism orientations, with the original spectral wavelength scales running along rows, and then along columns. Although these different orientations result in different nearby spectra, in both cases the Ly$\alpha$ line is independently detected at the same location and wavelength. The observed 2-D spectrograms are 1\farcs4 high,
while the simulated spectral images to the left are on approximately the same
scale.}
    \label{fig:jwst_Lya_sim}
\end{figure}

\begin{figure}[htbp]
    \centering
    \includegraphics[width=\textwidth]{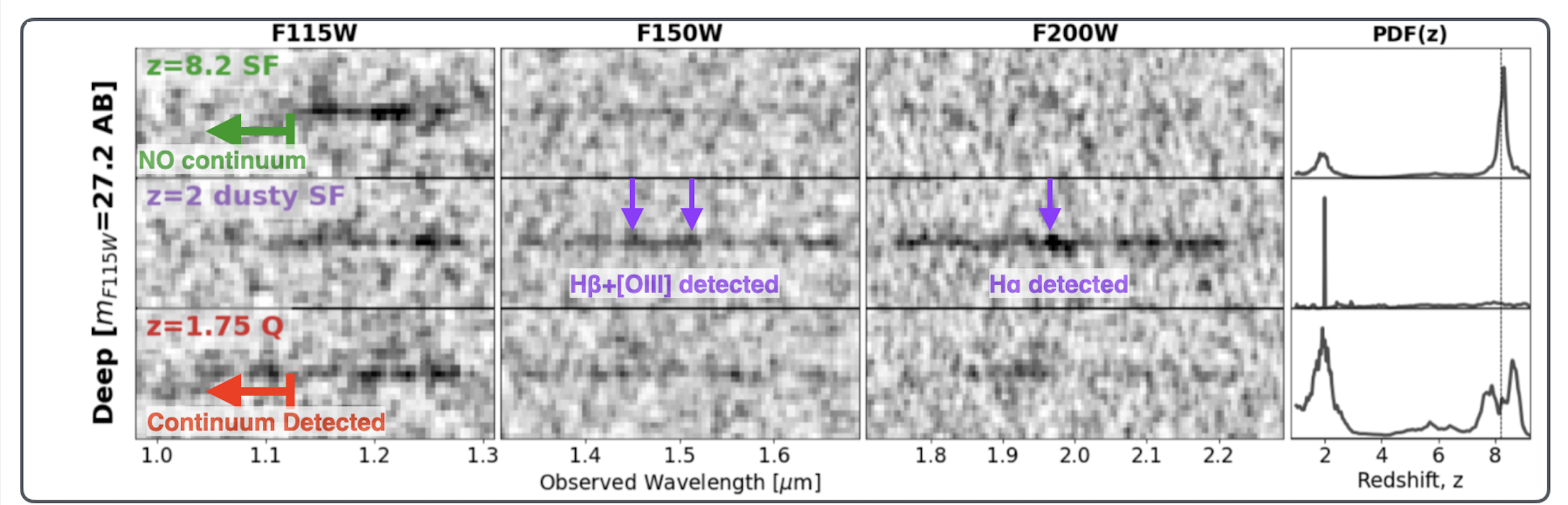}
    \caption{
     PASSAGE can distinguish between $z > 8$ Lyman break galaxies (LBG) and low-redshift contaminants. A simulated $z = 8.2$ LBG is shown in the top row, detected by its Lyman-alpha break (rest-frame 1216\AA) in the deep observations. The slitless 2D spectra for the F115W, F150W, F200W filters run from left to right. The spatial (vertical) scale is the same as in Figure \ref{fig:jwst_Lya_sim}. The last column shows the redshift probability distribution from template fitting. This simulation shows that PASSAGE correctly distinguishes between a $z \sim 8$ LBG and a $z = 2$ dusty star-forming galaxies (middle row) or a $z = 1.75$ passive galaxy (bottom row), which are prime contaminants. We caution that if less than all three wavelength regions are available, and no emission lines are detected, then the probability density functions (PDFs) for the high- and low-redshift solutions are not so unambiguously distinguished.} 
    \label{fig:jwst_Lya_Break_sim}
\end{figure}

\subsection{Feasibility of Spectroscopically Identifying and Measuring \texorpdfstring{Ly$\alpha$}{Lyman-alpha} lines in High Redshift Galaxies}
The total F115W grism exposures useful to search for faint Ly$\alpha$ lines and Lyman alpha forest breaks at $z\ge 8$, range from 8 ksec to more than 20 ksec in the deepest parallel visits. 
In the simulation shown in the left lower panel of Figure \ref{fig:jwst_Lya_sim}, a Ly$\alpha$ emission line is added to a Lyman break galaxy (LBG) spectrum. Even for these faint galaxies, the continuum break
and the line are confidently detected, for rest-frame line equivalent widths as small as 50\AA.

We have made a careful  search of our deepest PASSAGE fields for strong isolated emission lines in the 1.0--1.3 $\mu$m range. This special ``by-hand" analysis does not require that any object be initially detected in the broadband images. It has revealed 4 new 
Ly$\alpha$ emission lines at redshifts of z=7.7--9.5 \citep{Runnholm2025}. Their equivalent widths range from 40 to 100\AA. Figure \ref{fig:jwst_Lya_Break_sim} shows a simulated grism spectrum for a Lyman break galaxy emitting stellar continuum only. Although there is no Ly$\alpha$ emission line in the simulation, since there are many observed continuum pixels above and below the Lyman break (at 1.12$\micron$ = 1216\AA * 9.2), this sharp break is detected. And it is distinguishable from the gradual turn-downs seen in $z=2$ interloper galaxies, even if they lack any rest-frame optical emission lines. NIRISS slitless spectra have indeed detected Lyman breaks in 2 out of 3 faint $z\sim 8$ galaxies  \citep{Roberts22}, with magnitudes and integration times similar to these simulations. 
 
\section{Summary and Legacy Impact of PASSAGE} \label{sec:Summary}
\vskip0.1in
\textbf{Synergies with other programs and missions.} 
The JWST/NIRISS capability for slitless near-infrared spectroscopy in a large number of parallel fields has opened many new prospects for better understanding galaxy evolution. Our preliminary analysis of the first PASSAGE observations demonstrates that the performance of this mode is as good, or even better, than was anticipated before launch. We show recent NIRISS examples of major improvements in
sensitivity and extended long-wavelength coverage, to allow advances in the study of galaxies at higher redshifts and lower stellar masses than were previously attainable. We highlight progress in studying the star formation main sequence, the mass/metallicity correlation, and gas reddening.  PASSAGE complements ongoing deep JWST spectroscopy of selected fields and targets, by measuring galaxy spectra without pre-selection in 63 independent fields, to overcome uncertainties due to cosmic variance.

The pure parallels provide a public treasure trove of spectroscopy. This multiplies their future value, allowing for JWST follow-up, as well as 
complementary data for the Euclid and Roman missions on how to optimize observing strategies.
PASSAGE complements those missions by going much deeper in smaller areas of sky, thus conducting science that neither Euclid nor Roman missions can achieve.

The brighter (and rarer) objects discovered are particularly compelling for follow-up by JWST and other facilities.
For example, PASSAGE is discovering galaxies bright enough
to measure the key [\ion{O}{3}] $\lambda$4363 line, to test or improve the calibration of current metallicities at $z\ge 2$ (\cite{Gburek2019, Sanders2020}). Even if the high-redshift strong-line metallicity calibration does need to be revised, this project provided the essential suite of lines on which it is based.

\noindent \textbf{Legacy Value of PUBLIC fully processed datasets in the Archive.} 
High Level Science Products (HLSPs) are used up to 10 times more than individual calibrated data, and 
the archive of slitless spectroscopy is sure to provide an unprecedented legacy of lasting scientific value. 
\noindent We will deliver fully processed data and HLSPs to MAST for the astronomy community.

\vskip -0.1cm

\section{Acknowledgments} \label{acknowledgements}
We acknowledge support by NASA through grant JWST-GO-1571 (PI: Malkan). 
We are very grateful to staff members at Space Telescope Science Institute who worked hard to schedule pure parallel opportunities for this project, especially to our Program Coordinator Shelly Meyett. We thank the anonymous referee, for making extremely thorough readings and  providing many comments and suggestions which have improved this paper.
The JWST data presented in this article were obtained from the Mikulski Archive for Space Telescopes (MAST) at the Space Telescope Science Institute. The specific observations analyzed can be accessed via \dataset[doi: http://dx.doi.org/10.17909/6ca5-ba17]{http://dx.doi.org/10.17909/6ca5-ba17}.

XW is supported by the National Natural Science Foundation of China (grant 12373009), the CAS Project for Young Scientists in Basic Research Grant No. YSBR-062, the Fundamental Research Funds for the Central Universities, the Xiaomi Young Talents Program, and the science research grant from the China Manned Space Project. BV, AA, and PW acknowledge support from the INAF Large Grant 2022 “Extragalactic Surveys with JWST” (PI Pentericci) and from the European Union – NextGenerationEU RFF M4C2 1.1 PRIN 2022 project 2022ZSL4BL INSIGHT. PW and BV acknowledge support from the INAF Mini Grant `1.05.24.07.01 RSN1: Spatially Resolved Near-IR Emission of Intermediate-Redshift Jellyfish Galaxies' (PI Watson). AJB acknowledges funding from the “FirstGalaxies” Advanced Grant from the European Research Council (ERC) under the European Union’s Horizon 2020 research and innovation program (Grant agreement No. 789056). HA is supported by CNES, focused on the JWST mission and the Programme National Cosmology and Galaxies (PNCG) of CNRS/INSU with INP and IN2P3, co-funded by CEA and CNES.
MB acknowledges support from the ERC Grant FIRSTLIGHT and from the Slovenian national research agency ARRS through grants N1-0238 and P1-0188. MJH is fellow of the Knut \& Alice Wallenberg Foundation.
\bibliography{jwst_malkan}
\bibliographystyle{aasjournal}

\end{document}